\documentclass[aps,twocolumn,showpacs,superscriptaddress,floatfix,notitlepage,prx]{revtex4-2}

\usepackage{graphicx}%
\usepackage{multirow}%
\usepackage{amsmath,amssymb,amsfonts}%
\usepackage{amsthm}%
\usepackage{xcolor}%

\usepackage{hyperref}
\hypersetup{
colorlinks,
citecolor=blue,
filecolor=blue,
linkcolor=blue,
urlcolor=blue
}

\newcommand{\ket}[1]{\left|#1\right>}
\newcommand{\bra}[1]{\left<#1\right|}

\begin{document}

\title{Long-Time Error-Mitigating Simulation of Open Quantum Systems on Near Term Quantum Computers} 

 \author{Brian Rost}
 \email{bwr5@georgetown.edu}
 \affiliation{Department of Physics, Georgetown University; Washington DC.}

 \author{Lorenzo Del Re}
 \email{lorenzo.delre@gmail.com}
 \affiliation{Department of Physics, Georgetown University; Washington DC.}

 \author{Nathan Earnest}
 \affiliation{IBM Quantum, IBM T.J. Watson Research Center; Yorktown Heights, New York.}

 \author{Alexander F. Kemper}
 \email{akemper@ncsu.edu}
 \affiliation{Department of Physics, North Carolina State University; Raleigh, North Carolina.}

 \author{Barbara Jones}
 \affiliation{IBM Quantum, IBM Research Almaden; San Jose, California.}

 \author{James K. Freericks}
 \email{James.Freericks@georgetown.edu}
 \affiliation{Department of Physics, Georgetown University; Washington DC.}

    \begin{abstract}
       We study an open quantum system simulation on quantum hardware, which demonstrates
    robustness to hardware errors
    even with deep circuits containing up to two thousand entangling gates.
    We simulate two systems of electrons coupled to an infinite thermal bath: 1)  a system of dissipative free electrons
    in a driving electric field; and 2) the thermalization of two interacting electrons in a single orbital in a magnetic field---the Hubbard atom. These problems are solved using IBM quantum computers, showing no signs of decreasing fidelity at long times. Our results demonstrate that algorithms for simulating open quantum systems
    are able to far outperform similarly complex non-dissipative algorithms on noisy hardware. Our two examples show promise that the driven-dissipative quantum many-body problem can eventually be solved on quantum computers.
    \color{black}
    \end{abstract}


\maketitle

\section{Introduction}

Quantum systems in nature are inherently open --- they inevitably interact with an environment, except in extreme cases of perfectly isolated particles. As a prototypical example, we consider condensed-matter systems, where the electrons that exhibit interesting emergent phenomena (such as superconductivity) and their strongly coupled lattice vibrations
co-exist with
a hard-to-characterize environment that has contributions from, for example,  
remaining lattice vibrations and spurious external electromagnetic fields.
Nevertheless, we observe emergent
quantum phenomena in nature, which indicates that the presence of these (typically dissipative) effects does not destroy the emergence of complex quantum physics. This is because the open quantum system has fixed points\cite{albert2014symmetries,buvca2019non} that continue to exhibit the emergent phenomena. Thus, we can theoretically ignore the open-ness, and focus on the emergent phase within an approximate closed-system description. 

The situation gets more complicated when driving fields are introduced. Now, the dissipative effects are crucial for determining the fixed point(s),  called non-equilibrium steady state(s) (NESS), and thus a full open system  needs to be solved.  This requires the use of a density matrix formalism, which further exacerbates the ``curse of dimensionality'' that already plagues pure state simulation. However, this
is also a research frontier in many different disciplines: 
(i) in quantum condensed-matter physics, pump-probe experiments drive a system into nonequilibrium and watch how it evolves~\cite{pump-probe-review}; 
(ii) in chemistry, driving and dissipation play important roles in photochemistry~\cite{thyrhaug2018identification} and in cavity-enhanced reactions~\cite{cavity}; 
(iii) in nuclear and high-energy physics, collisions of heavy nuclei induce nonequilibrium dynamics~\cite{high-energy}; 
(iv) in quantum optics, cavity QED studies dressed atoms in a cavity and how they reach a steady state\cite{haroche2013nobel}; and many more. In some cases, the system is driven to new (meta)stable non-equilibrium phase that cannot be produced any other way\cite{basov2017towards}. In this work, we are interested in both steady states that do not change, and limit cycle states, which have periodically repeating ``steady states.''

From an applications perspective, many devices rely upon the nonlinear response of the system to external fields---the switchable nonlinear current-voltage curve of a transistor is a classic example. As quantum materials are sought for use in novel device applications, understanding how they respond to fields in the presence of dissipation is critical for engineering these devices. Hence, driven dissipative systems also have an almost ubiquitous appearance through many fields of applied science.
%

An emergent technique for modeling quantum systems is to use quantum computers. These are naturally open systems by
themselves:
the qubits interact with an uncontrollable environment.
They
undergo $\tau_1$ and $\tau_2$ decay, and
the fixed point of their evolution is 
typically the $|0\rangle\langle 0 |$ state, i.e. the individual qubit's ground state. Indeed, much of the progress on today's quantum hardware is limited by the decoherence of the individual qubits\cite{arute2020observation}.
Efforts are underway to improve this situation, but with an eye towards closed quantum simulation, which is sensitive to the environment, and which does not typically exhibit fixed points in the dynamics. The situation improves when simulating open systems on a quantum 
computer\cite{del2020driven,zanardi2015geometry,schirmer2010stabilizing,jaschke2019thermalization,hu2020quantum,rost2020simulation}, which is a topic that has garnered some interest recently\cite{tacchino2020quantum,de2021quantum,ramusat2021quantum,schlimgen2022quantum,hu2022general,wang2023simulating,mi2023stable}.

Here we show that modeling a driven-dissipative system on a quantum computer is intrinsically stable on the hardware, even in the presence of noise and decoherence.
The stability happens because the dynamics maps all initial states that are not protected in some way onto the fixed point of the evolution, which is a periodic NESS.  This naturally includes states that have undergone some perturbation by the noise in the hardware, and in this way, errors incurred in one time step are corrected by the inherent dynamics in the next step.  The fixed point is determined by a combination of the quantum channel corresponding to the evolution of the system under study, and the intrinsic quantum channel of the noisy quantum computer. The quantum-computer channel shifts the fixed point slightly, which is borne out in our results, and we show that the clean limit can be obtained via extrapolation of the noisy results.
Because of this inherent stability and because these problems are challenging to solve on conventional computers, the driven-dissipative problem is an excellent application for near-term quantum computing. 

We illustrate how dissipative quantum simulation can be performed on IBM's quantum computers by considering two problems derived from the Hubbard model in two distinct limits: 1) 
a noninteracting system of lattice electrons at finite electric field and temperature; and 2) an interacting single-site two electron system in a magnetic field that thermalizes---the Hubbard atom. As discussed below, while the free fermionic problem is trivially solvable in a closed system, the presence of dissipation adds significant complications that inhibit a simple solution of the (polynomially-sized) 1-body Hamiltonian.

The key new feature that allows for open quantum system simulation on near term hardware is the mid-circuit reset gate.
There are two ways to use this to simulate dissipation into a reservoir or environment --- one way simulates the full system plus reservoir and uses mid-circuit resets to dissipate energy from the qubits that represent the reservoir, while the second way integrates out the reservoir and implements more complicated dynamics involving superoperators. In the first case, one works with a larger Hilbert space corresponding to the system plus reservoir, but the dynamics are unitary time evolution, followed by reset operations on the reservoir qubits which remove energy and create mixed states~\cite{mi2023stable}. In the second case, one works with the smaller Hilbert space of the system only (and hence can incorporate infinite-sized reservoirs), but the time dynamics is that of a non-unitary quantum channel, which is more complicated to implement. In this work, we follow the second approach.

\section{Results}

\subsection{Approach}
Although many quantum algorithms are known for the simulation of closed quantum systems, fewer studies have considered the simulation of open quantum systems despite their rich and interesting behavior~\cite{breuer2002theory}. Current approaches 
using inherent qubit decoherence~\cite{sommer2021,rost2020simulation,tseng2000quantum}, direct simulation of an environment~\cite{terhal2000problem,wang2011quantum,su2020quantum}, implementing Kraus maps/Lindblad operators~\cite{cleve2016efficient,hu2020,hu2022,head2021capturing,childs2016efficient,kliesch2011dissipative}, variational techniques~\cite{haug2020generalized,yoshioka2020variational}, and more~\cite{kamakari2021,metcalf2020engineered}. Since Barreiro et al. first demonstrated their open-system quantum simulator~\cite{Barreiro_2011}, current early-stage dissipative simulations of quantum systems in the areas of quantum chemistry and physics~\cite{kamakari2021,hu2020,rost2020simulation,tornow2022non,de2021quantum,schlimgen2022quantum,hu2022general,wang2023simulating,mi2023stable} have been completed.
Here, we implement
a Trotterized driven-dissipative map, which
involves the application of Kraus operators
at each time step\cite{hu2020,hu2022,head2021capturing,childs2016efficient,kliesch2011dissipative}.
On a quantum computer where the natural operations are unitary, we implement the non-unitary gates by coupling to an ancilla qubit which is subsequently reset; this produces a non-unitary channel resulting
in an eventual (mixed) state.

For our models, we choose two endpoints of the interacting Hubbard model.  The first endpoint (noninteracting lattice electrons) is a system of free fermions in the presence of infinitely sized dissipative bath; the second endpoint (interacting electrons on a single site) is the Hubbard atom.  Note that, in stark contrast to the closed system case, where free fermionic models are exactly solvable, the addition of dissipation renders the model to be non-trivial.
An exact solution for a metal with a constant density of states does exist\cite{han2013solution}, but it is not
trivial to obtain; it requires
using the Keldysh Green’s function formalism to complete it.
The choice of an (integrated out) infinite bath is critical to avoid finite size effect oscillations that arise from directly simulating the reservoir\cite{mi2023stable}, which
inhibit the stability of the fixed point.

We underscore that all of the steps that need to be implemented are the same for a non-interacting system as for an interacting one. The main differences are that (i) by working in the single particle basis, the Hilbert space is much smaller for the non-interacting system and (ii) that the computational basis is the appropriate basis for the Kraus operators. Interacting systems are primarily complicated by the nontrivial basis needed for the Kraus operators; when exactly integrating out a reservoir the number of Kraus operators is given by the square of the Hilbert space dimension, and they map energy eigenstates to energy eigenstates, making this "straightforward" approach impossible to carry out on a quantum computer, except for very small systems. 
Thus, although our demonstration does not cover
the full range of systems,
the approach can be extended more broadly once an efficient representation for the Kraus operators is found, or other more efficient algorithms are discovered which allow single time steps to be carried out with high enough fidelity.

The concrete details for how we determine the algorithms here are given in the Supplementary Information. The specific systems we study have the benefit that the computational basis is a natural basis for the Kraus operators, which allows their implementation to be quite efficient. In addition, many of the Kraus operators are not needed to achieve the desired time evolution.
This is because we are not simulating a specific reservoir, but instead are simulating a generic one. In the case of the Hubbard atom, because we are going to a thermal state, and this state can be universally prepared by any thermal reservoir, we can greatly reduce the number of Kraus operators needed. In order to be be able to carry out this work, the most efficient implementation of the open system is required in order to have sufficient fidelity for each Trotter step.

\subsection{Infinite 1D chain of driven-dissipative fermions}

\begin{figure*}[t]
	\centering
	\includegraphics[width=0.99\textwidth]{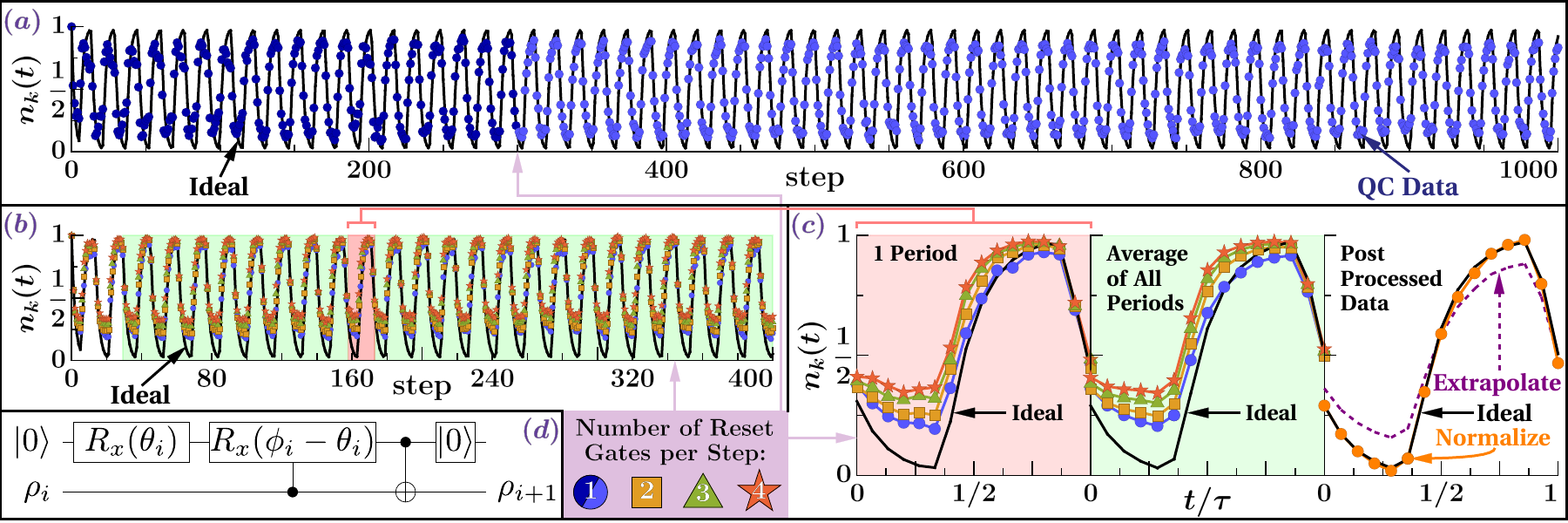}
	\caption[Electron density vs. time]{(Color online) Electron density vs. time. \textbf{(a)} 1000 steps of time evolution on \texttt{ibmq\char`_mumbai} using one reset gate per Trotter step. Reported data has been corrected for measurement errors. Statistical errors are at about the 1.5\% level, smaller than the size of the plotting symbol (so error bars are suppressed). Different shades of blue represent different sets of qubits while the solid black line is the ideal result of running the circuit. \textbf{(b)} 400 steps of time evolution on \texttt{ibmq\char`_boeblingen} using one to four reset gates per Trotter step. Reported data has been corrected for measurement errors. \textbf{(c)} Left panel shows a zoom in on one period of the data shown in (b). Middle panel shows one period of data obtained by averaging together all periods of steady-state data (step $>30$). Right panel shows post-processed averaged data as described in the technical details section. \textbf{(d)} Quantum circuit for one Trotter step of time evolution. Here, $\theta_i=2 \sin ^{-1}\sqrt{\frac{2\Gamma  \Delta t}{e^{\beta  \varepsilon_i}+1}}$ and $\phi_i=2 \sin ^{-1}\sqrt{\frac{2\Gamma  \Delta t}{e^{-\beta  \varepsilon_i}+1}}$ where $\varepsilon_i=-2\cos(k+\Omega i\Delta t)$ is the dispersion relation at step $i$ (at time $t=i\Delta t$), and we use $\Gamma=0.1$.
	}
	\label{fig:nk}
\end{figure*}

First, we examine the free fermion limit of a Hubbard model in the presence of a driving field
and dissipative coupling to a bath.
Electrons freely move on an infinite one-dimensional lattice via nearest-neighbor hopping (used as our energy unit), with semi-infinite electronic thermal reservoirs (taken in the wide-band limit) attached to each lattice site. A DC electric field of strength $E$ is applied by employing a linearly varying Peierls phase $\phi(t)=\Omega t$  ($\Omega=eEa$, with $e$ the electric charge and $a$ the lattice spacing (we set $\hbar$ and $c$ to one) in the lattice hopping parameter.
Each lattice site is coupled to a reservoir
by hopping term to one end of a semi-infinite reservoir;
all properties of this coupling are summarized in a parameter $\Gamma$ given by the square of the lattice-reservoir hopping multiplied by the density of states of each reservoir. 

The infinite system is diagonalized by Fourier transforming to momentum space. In this fashion,  the dynamics of the infinite lattice is addressed by solving many independent single-qubit systems, each of which depends on the specific value of the crystalline momentum $k$; in the steady state, we construct the results for all momenta from the results for any single momentum by invoking gauge invariance. The key point is that a constant DC electric field shifts the momentum linearly in time (when we employ the Peierls substitution). This means that the results from different momentum values can be found by simply shifting the time axis of the results for one momentum 
(further details are presented in the Supplementary Information).
Hence, the properties of the infinite lattice can be determined by the results for the time trace of just one momentum. The time and momentum dependent Kraus operators (for the time step of duration $\Delta t$) are given by:
\begin{subequations}
\begin{align}\label{eq:kraus}
    K_0 &= 
     d^\dag_k d^{\,}_k \sqrt{1-2\,\Gamma n_F[-\epsilon_k(t)]\Delta t}\,e^{-i\epsilon_k(t) \Delta t} \\ & + d^{\,}_kd^\dag_k \sqrt{1-2\Gamma n_F[\epsilon_k(t)]\Delta t} \\
    K_1 &= \sqrt{2\Gamma\,\Delta  t\,n_F[\epsilon_k(t)]}\,d^\dag_k \\
    K_2 &= \sqrt{2\Gamma \Delta t\,n_F[-\epsilon_k(t)]}d_k^{\,},
\end{align}
\end{subequations}
where $d^\dag_k$ is the creation operator of a lattice electron with momentum $k$, $\epsilon_k(t)$ is the time dependent dispersion relation, 
$\Gamma$ is the strength of the coupling to the reservoir (here set to 0.1),
and $n_F(x)=1/(1+e^{\beta x})$ is the Fermi-Dirac distribution at the temperature of the reservoirs (here set to $\beta=\frac{1}{k_BT}=5$).

From the Kraus map given in Eq.~(1) the quantum circuit can be constructed~\cite{del2020driven} and appears in Fig.~\ref{fig:nk}(d).
The $|0\rangle$ gate in the circuit is a reset operation which ideally sets the qubit to the $|0\rangle$ state. However, this operation is not perfect on current hardware and the reset fidelity can be improved by applying it to a qubit multiple times in succession (at the cost of additional amplitude damping on the remaining qubits, because the reset operation takes by far the longest time to run compared to the rest of the circuit). We run our circuit using one to four reset gates per Trotter step. This data is used to build an effective model of the error combining these two effects, which we use to extrapolate away some of the errors. Details are given in 
the Supplementary Information.

Figure~\ref{fig:nk}(a) plots the results of running the circuit shown in \ref{fig:nk}(d) 
with a single reset gate per Trotter step. The first 300 steps were run on one set of qubits and the remaining 700 on a second set. These runs are robust both with respect to the choice of qubits and with respect to the total time of the run. The circuit for the 1000\textsuperscript{th} step required 2000 CNOT gates and yet the data show no sign of a decaying signal.  Note that the transients have died off after about 30 Trotter steps. In panel (b), the raw data with one to four resets is shown for up to 400 Trotter steps. After processing the data via extrapolating to the zero-reset-time limit and then correcting the result by shifting and stretching as described in the SI, one can see that the post-processed data (c) agrees with the exact results to high precision. 
This simulation clearly shows the stability of driven-dissipative circuits on near-term quantum
computers, producing accurate time evolution far longer than is currently possible with the quantum simulation of coherent dynamics of electrons~\cite{arute2020observation}.
\begin{figure}[ht]
	\centering
	\includegraphics[width=\columnwidth]{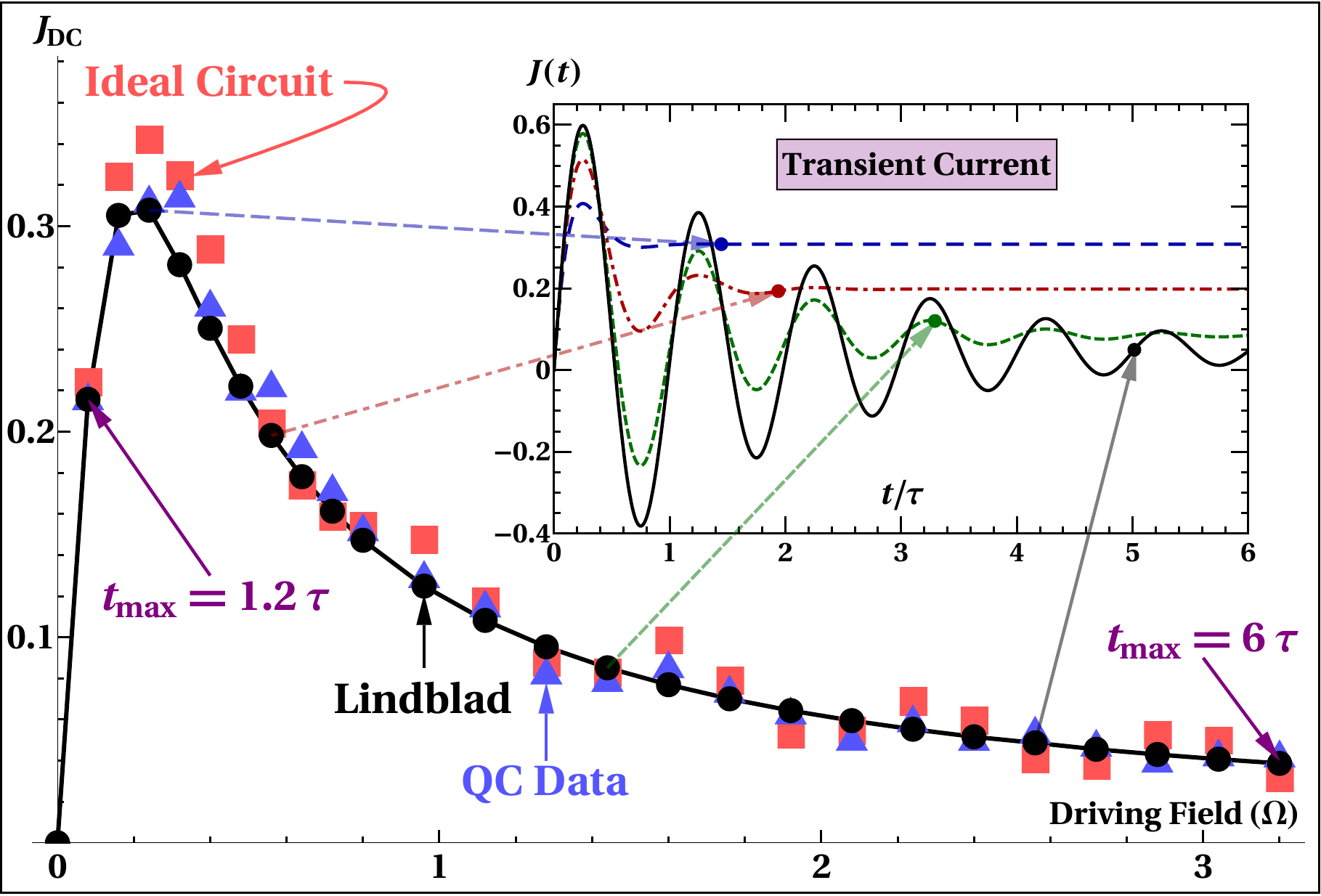}
	\caption{(Color online) DC current versus electric field strength. Comparison of the DC response at a given field strength as computed (see technical details section) using: (i) the Lindblad master equation (black circles), (ii) the ideal circuit (red squares), and (iii) the measured data from \texttt{ibmq\char`_boeblingen} (blue triangles). All circuits for the current were run for 50 Trotter steps each due to total quantum computer time available, leading to a trade-off between Trotter error and convergence error. $\Delta t$ was chosen empirically such that $\Delta t/\tau$ is linear in $\Omega$. Convergence to the steady-state current is shown in the inset with the solid dots representing $t_\text{max}$ for that run.}
	\label{fig:current}
\end{figure}

Figure~\ref{fig:current} shows the steady-state DC current response (averaged over one oscillation) of the system to an applied electric field, and compared against the ideal circuit and theoretical results from Lindblad techniques, as described in Ref.~\cite{del2020driven}. 
We can see an interesting result has developed. The Bloch oscillations that are characteristic of free electrons in an electric field in the presence of a heat bath in our model give rise to a net DC current. As a function of external field, the current first increases as expected because more energy is put into the system, but soon reaches a maximum and then decays. The reason for this is illustrated by the inset of Fig.~\ref{fig:current}, which shows that as the electric field is increased, there are more oscillations, and integrated, these cancel out. One can see that the maximum current is characterized by minimal transient  oscillations.

That the the transient region extends to longer times as the electric field is increased has practical ramifications as well, requiring the simulation to run further in time. Hence, the calculation requires a trade-off between Trotter error (large time step size) and convergence error (steady state not yet reached). To minimize the total error, we choose $\Delta t$ empirically with $\Delta t/\tau\approx 0.022+0.031 \Omega$ (here, $\tau=2\pi/\Omega$). At large driving fields it is clear that we have not yet reached the steady state and oscillations in the DC current data are observed even in the ideal case. At small driving field $\Delta t=\mathcal{O}(1/\Omega)$ and we incur large Trotter errors. Despite these limitations, the quantum computer results match both the ideal circuit and the theoretical results fairly accurately across the entire range of simulated field strengths.
\subsection{Interacting orbital with magnetic field and baths}
\begin{figure*}[t!]
	\centering
	\includegraphics[width=0.99\textwidth]{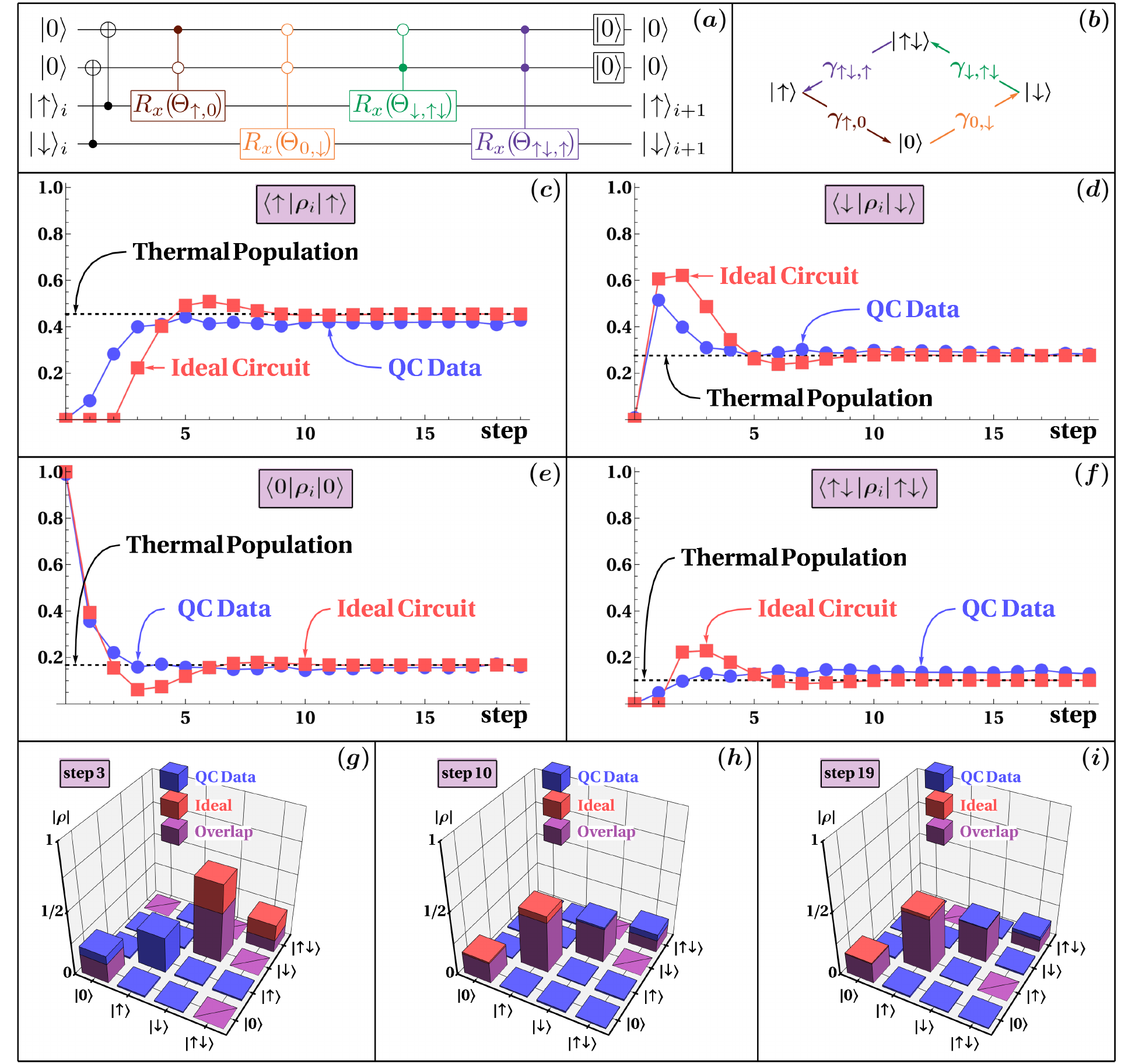}
	\caption{(Color online) Thermalized Hubbard model in the single-site limit, with magnetic field. \textbf{(a)} Circuit for a single Trotter step of time evolution. The rotation angles $\Theta$ are given in terms of transition probabilities $\gamma$, which are derived from the detailed balance condition as depicted in \textbf{(b)} and described in SI. $\Theta_{i,j}=2\sin^{-1}(\sqrt{\gamma_{i,j}})$ and $\gamma_{i,j}\propto e^{\beta \varepsilon_i}$, where $\varepsilon_i$ is the energy of state $|i\rangle$. \textbf{(c)--(f)} Results showing the populations of each of the four possible occupation states versus Trotter step for the ideal case (red squares), measurement-error mitigated data from \texttt{ibmq\char`_mumbai} (blue circles), and the theoretical thermal population (dashed black line). \textbf{(g)--(i)} Full tomography of the density matrix at selected time steps.
	} 
	\label{fig:hubbard}
\end{figure*}

The second problem we simulate is the atomic limit of the dissipative Hubbard model in an external magnetic field $B$,  
where
the bath may be either hotter or colder than the system. This system with maximum two-electrons is a strongly correlated electron problem, which can be simulated with four qubits. The energy cost of double occupancy is given by $U$ and the magnetic field shifts the energy levels of the single-spin states up or down with an energy splitting given by $B$, measured in units of the Bohr magneton. The Hamiltonian is
\begin{equation}
\mathcal{H}=Un_{\uparrow}n_{\downarrow}-\frac{\mu}{2} (n_{\uparrow}+n_{\downarrow})-\frac{B}{2}(n_{\uparrow}-n_{\downarrow})
\end{equation}
where $U$ is the on-site interaction strength, $\mu$ is the chemical potential, $B$ is the magnetic field in units of energy, and $n_{\alpha}$ is the occupation number operator for state $\ket{\alpha}$.
$B$ is chosen to have a sign such that energy is lowered by the occupation of a single up electron.

We build our dissipative circuit by using Kraus operators that induce transitions between computational basis states, which are the energy eigenstates. These transitions are implemented by mapping the system state to an ancilla register and rotating the system qubits controlled on the ancilla~\cite{barreiro2011open,del2020driven} (shown in Fig.~\ref{fig:hubbard}(a)); the approach uses just the transitions that correspond to a cycle through the states  (depicted in Fig.~\ref{fig:hubbard}(b)) Details are given in the Supplementary Information. This circuit is iterated $n=19$ times to simulate $n$ Trotter steps of time evolution.

Figure~\ref{fig:hubbard} shows the transient evolution of the Hubbard atom from the vacuum state (with no electrons) to a thermal state with filling $n\approx 0.83$ at a temperature $T=U/2$ in an external magnetic field $B=U/4$ using one reset per Trotter step. We note the effect of finite temperature on the results. Indeed the state parallel to the magnetic field (c) has the highest population at steady state, but it 
remains below $1/2$.
 The next highest population is the state with one down electron (d), which although paying a cost of being opposite the magnetic field, still is not causing a Coulomb interaction $U$. Note the relatively large fluctuation to large population for the one down electron as the filling is ramping up from vacuum state, partly a result of noise, but mainly because (see the directionality of the arrows in Figure~\ref{fig:hubbard}(b)) this state is filled first from the vacuum. The state with no electrons (e) is of course the initial condition, so it shows occupancy 1 at step 0, but soon drops to low occupancy, but not so low as the state with two electrons (f), which suffers the energy cost $U$.

These data are only post-processed for measurement-error corrections. The transient data lie close to the ideal circuit, but show deviations due to intrinsic errors in the hardware. Nevertheless, the steady state is reproduced accurately for the four different populations of the thermal state (c-f) and for the thermal final density matrix (g-i). This further exemplifies the robustness of these types of algorithms to noise and how errors in early time steps are largely corrected in subsequent steps.

\section{Discussion}

In this work, we have demonstrated that the capabilities of near-term quantum hardware
for simulating open quantum systems are greater than one might expect based on the 
state of the art results in simulating closed systems.  By making use of the newly developed
reset gates and mid-circuit measurements, we have achieved dynamical simulations,
with 1000 Trotter steps using circuits of up to 2000 CNOT gates with minimal
signal decay.  We interpret this success by viewing the dissipative dynamics as a dynamical
map with a fixed point;  the hardware noise is not sufficient to overcome the tendency of the map
to drive the systems towards their respective fixed points. It may perturb the fixed point, but does not significantly modify its character, which is what makes the process robust.

For the models discussed here, we observe that the dynamical map of the evolution has a unique
fixed point, and that the fixed point is only mildly affected by the hardware noise and the Trotter time-step infidelity. This is evidenced by the
relatively good agreement between the ideal circuit and the QC data shown in Figs~\ref{fig:nk} and \ref{fig:hubbard}. In general, however, it may be that in other scenarios the fixed point is sufficiently sensitive to the particular effects
of the noise that the resulting evolution retains very little of the ideal circuit. 
Similarly, we expect that as the noise gets worse, for example when scaling to larger systems, that the
modification of the dynamical map becomes more severe.
We reserve these questions 
for future investigation. 
However, we do note a simple empirical result;
if {\emph{a single}} Trotter step has sufficiently high fidelity --- i.e. the map of the open system combined with the intrinsic decoherence of the quantum computer and the infidelity of the Trotter step yields a fixed point that is close enough to the fixed point of the original open system ---  then the simulation is successful and can be accurately continued to long times. This result is similar in nature to the well-known result for error correction that the fidelity of the individual qubits must be sufficiently high in order for it to be possible to carry out error correction. The analog here is the sufficiently high fidelity of a single Trotter step.

These conclusions bring up an important question---under what circumstances will the perturbation of the fixed point be small and controllable so the quantum simulation will be successful? We are not able to answer this question here, and we believe it is an important question for the community to address given the results we have found, which indicate rather broad robustness. In order for these types of problems to be solved on quantum computers knowing which systems will be successful and which ones will fail is obviously an important area for future work.

The community is searching for nontrivial problems that can be solved on quantum computers that show an advantage over classical computers. While our work has not yet achieved this goal, it does show that this class of problems 
is a potential path towards
 achieving this goal. Aside from our demonstration here, another reason is that the non-equilibrium many body problem does not have efficient algorithms to solve it on classical computers. The algorithms that do exist usually are restricted to either very small size systems or to short times. Here, we show that on a quantum computer, there may not be a short-time restriction, and this is a very important result for using them to simulate nontrivial systems.

\section{Methods}
All data were taken using 
superconducting
quantum computers made available by IBM, either \texttt{ibmq\char`_mumbai} (Fig.~\ref{fig:nk}(a) and Fig.~\ref{fig:hubbard}) or \texttt{ibmq\char`_boeblingen} (Fig.~\ref{fig:nk}(b,c) and Fig.~\ref{fig:current}). \texttt{ibmq\char`_boeblingen} is a 20-qubit quantum device and \texttt{ibmq\char`_mumbai} is 27-qubit device. 
These two devices were chosen because they were among the first with
reset gates available.
The reported error rates were used to select sets ($\sim$5-10) of candidate qubits on which a limited number of Trotter steps were run. Results from these were then used to select the final qubits on which the full job(s) would be run. Raw shot counts were processed using Qiskit Ignis'~\cite{qiskit2024} built in measurement error mitigation protocol. This prepares and immediately measures each computational basis state giving a confusion matrix, which is inverted and applied to the raw shot counts to yield the mitigated shot counts.  For each data point shown in the figures, we used 1,000 shots.

This work pushed current near-term quantum computers to their limits. This created some unique issues when trying to run these extremely deep circuits. 
We encountered buffer overflow errors after running a large number of our larger circuits indicating we had overflowed the device's capacity to record more measurement data. This is why we limited our DC current data to 50 Trotter steps per unique set of parameter values. We avoided this limit when taking data for Fig.~\ref{fig:nk}(b) by breaking our jobs into smaller chunks, but this meant long queue times, which would have been prohibitive to get the data in Fig.~\ref{fig:current}. The data in Fig.~\ref{fig:nk}(a) were obtained through exclusive access to \texttt{ibmq\char`_mumbai}, which allowed us to break our circuits into individual jobs without worrying about queue times. Nevertheless, we still overflowed the buffer around step 300, forcing us to change to a new pair of qubits with a fresh buffer. Finally, we began to exceed the limits of the system generating the driving microwave pulses at around 1000 Trotter steps, which is why this is the upper limit for Fig.~\ref{fig:nk}(a).  

For the full tomography results in Fig. \ref{fig:hubbard}, we made use of the ``state\_tomography\_circuits'' function and ``StateTomographyFitter'' class built into Qiskit Ignis.  The ``state\_tomography\_circuits'' creates a list of $3^n$ circuits which carries out the desired quantum circuit and then measures in the $X$, $Y$, and $Z$ bases. These results are then fed into the ``StateTomographyFitter.fit'' function in order to reconstruct the full quantum state.

\section*{Acknowledgements}
We acknowledge use of the IBM Quantum devices for this paper.
We acknowledge financial support
from the U.S. Department
of Energy, Office of Science, Basic Energy Sciences, Division of Materials Sciences and Engineering.
Initial submission and project execution were performed under Grant No.
DE-SC0019469; the resubmission under
Grant no. DE-SC0023231.
We also acknowledge funding from the McDevitt bequest at Georgetown (JKF)
National Science Foundation QISE-NET Award No. DMR-1747426 (BR)
Aspen Center for Physics through National Science Foundation Grant No. PHY-1607611 (BJ).

\section*{Competing interests}
The authors declare that they have no competing interests. 

\section*{Data availability}

The data for the figures are available at\\
\href{https://datadryad.org/stash/share/b4V31s8N57c6Esik7HTlu0jvAlxjywEDja54yDntUdA}{https://datadryad.org/stash/
share/
\newline{}
b4V31s8N57c6Esik7HTlu0jvAlxjywEDja54yDntUdA}.

\bibliography{scibib}

\begin{thebibliography}{43}%
\makeatletter
\providecommand \@ifxundefined [1]{%
 \@ifx{#1\undefined}
}%
\providecommand \@ifnum [1]{%
 \ifnum #1\expandafter \@firstoftwo
 \else \expandafter \@secondoftwo
 \fi
}%
\providecommand \@ifx [1]{%
 \ifx #1\expandafter \@firstoftwo
 \else \expandafter \@secondoftwo
 \fi
}%
\providecommand \natexlab [1]{#1}%
\providecommand \enquote  [1]{``#1''}%
\providecommand \bibnamefont  [1]{#1}%
\providecommand \bibfnamefont [1]{#1}%
\providecommand \citenamefont [1]{#1}%
\providecommand \href@noop [0]{\@secondoftwo}%
\providecommand \href [0]{\begingroup \@sanitize@url \@href}%
\providecommand \@href[1]{\@@startlink{#1}\@@href}%
\providecommand \@@href[1]{\endgroup#1\@@endlink}%
\providecommand \@sanitize@url [0]{\catcode `\\12\catcode `\$12\catcode
  `\&12\catcode `\#12\catcode `\^12\catcode `\_12\catcode `\%12\relax}%
\providecommand \@@startlink[1]{}%
\providecommand \@@endlink[0]{}%
\providecommand \url  [0]{\begingroup\@sanitize@url \@url }%
\providecommand \@url [1]{\endgroup\@href {#1}{\urlprefix }}%
\providecommand \urlprefix  [0]{URL }%
\providecommand \Eprint [0]{\href }%
\providecommand \doibase [0]{https://doi.org/}%
\providecommand \selectlanguage [0]{\@gobble}%
\providecommand \bibinfo  [0]{\@secondoftwo}%
\providecommand \bibfield  [0]{\@secondoftwo}%
\providecommand \translation [1]{[#1]}%
\providecommand \BibitemOpen [0]{}%
\providecommand \bibitemStop [0]{}%
\providecommand \bibitemNoStop [0]{.\EOS\space}%
\providecommand \EOS [0]{\spacefactor3000\relax}%
\providecommand \BibitemShut  [1]{\csname bibitem#1\endcsname}%
\let\auto@bib@innerbib\@empty
\bibitem [{\citenamefont {Albert}\ and\ \citenamefont
  {Jiang}(2014)}]{albert2014symmetries}%
  \BibitemOpen
  \bibfield  {author} {\bibinfo {author} {\bibfnamefont {V.~V.}\ \bibnamefont
  {Albert}}\ and\ \bibinfo {author} {\bibfnamefont {L.}~\bibnamefont {Jiang}},\
  }\bibfield  {title} {\bibinfo {title} {Symmetries and conserved quantities in
  lindblad master equations},\ }\href@noop {} {\bibfield  {journal} {\bibinfo
  {journal} {Physical Review A}\ }\textbf {\bibinfo {volume} {89}},\ \bibinfo
  {pages} {022118} (\bibinfo {year} {2014})}\BibitemShut {NoStop}%
\bibitem [{\citenamefont {Bu{\v{c}}a}\ \emph {et~al.}(2019)\citenamefont
  {Bu{\v{c}}a}, \citenamefont {Tindall},\ and\ \citenamefont
  {Jaksch}}]{buvca2019non}%
  \BibitemOpen
  \bibfield  {author} {\bibinfo {author} {\bibfnamefont {B.}~\bibnamefont
  {Bu{\v{c}}a}}, \bibinfo {author} {\bibfnamefont {J.}~\bibnamefont
  {Tindall}},\ and\ \bibinfo {author} {\bibfnamefont {D.}~\bibnamefont
  {Jaksch}},\ }\bibfield  {title} {\bibinfo {title} {Non-stationary coherent
  quantum many-body dynamics through dissipation},\ }\href@noop {} {\bibfield
  {journal} {\bibinfo  {journal} {Nature Communications}\ }\textbf {\bibinfo
  {volume} {10}},\ \bibinfo {pages} {1730} (\bibinfo {year}
  {2019})}\BibitemShut {NoStop}%
\bibitem [{\citenamefont {Giannetti}\ \emph {et~al.}(2016)\citenamefont
  {Giannetti}, \citenamefont {Capone}, \citenamefont {Fausti}, \citenamefont
  {Fabrizio}, \citenamefont {Parmigiani},\ and\ \citenamefont
  {Mihailovic}}]{pump-probe-review}%
  \BibitemOpen
  \bibfield  {author} {\bibinfo {author} {\bibfnamefont {C.}~\bibnamefont
  {Giannetti}}, \bibinfo {author} {\bibfnamefont {M.}~\bibnamefont {Capone}},
  \bibinfo {author} {\bibfnamefont {D.}~\bibnamefont {Fausti}}, \bibinfo
  {author} {\bibfnamefont {M.}~\bibnamefont {Fabrizio}}, \bibinfo {author}
  {\bibfnamefont {F.}~\bibnamefont {Parmigiani}},\ and\ \bibinfo {author}
  {\bibfnamefont {D.}~\bibnamefont {Mihailovic}},\ }\bibfield  {title}
  {\bibinfo {title} {Ultrafast optical spectroscopy of strongly correlated
  materials and high-temperature superconductors: a non-equilibrium approach},\
  }\href@noop {} {\bibfield  {journal} {\bibinfo  {journal} {Adv. Phys.}\
  }\textbf {\bibinfo {volume} {65}},\ \bibinfo {pages} {58} (\bibinfo {year}
  {2016})}\BibitemShut {NoStop}%
\bibitem [{\citenamefont {Thyrhaug}\ \emph {et~al.}(2018)\citenamefont
  {Thyrhaug}, \citenamefont {Tempelaar}, \citenamefont {Alcocer}, \citenamefont
  {{\v{Z}}{\'\i}dek}, \citenamefont {B{\'\i}na}, \citenamefont {Knoester},
  \citenamefont {Jansen},\ and\ \citenamefont
  {Zigmantas}}]{thyrhaug2018identification}%
  \BibitemOpen
  \bibfield  {author} {\bibinfo {author} {\bibfnamefont {E.}~\bibnamefont
  {Thyrhaug}}, \bibinfo {author} {\bibfnamefont {R.}~\bibnamefont {Tempelaar}},
  \bibinfo {author} {\bibfnamefont {M.~J.}\ \bibnamefont {Alcocer}}, \bibinfo
  {author} {\bibfnamefont {K.}~\bibnamefont {{\v{Z}}{\'\i}dek}}, \bibinfo
  {author} {\bibfnamefont {D.}~\bibnamefont {B{\'\i}na}}, \bibinfo {author}
  {\bibfnamefont {J.}~\bibnamefont {Knoester}}, \bibinfo {author}
  {\bibfnamefont {T.~L.}\ \bibnamefont {Jansen}},\ and\ \bibinfo {author}
  {\bibfnamefont {D.}~\bibnamefont {Zigmantas}},\ }\bibfield  {title} {\bibinfo
  {title} {Identification and characterization of diverse coherences in the
  fenna--matthews--olson complex},\ }\href@noop {} {\bibfield  {journal}
  {\bibinfo  {journal} {Nature chemistry}\ }\textbf {\bibinfo {volume} {10}},\
  \bibinfo {pages} {780} (\bibinfo {year} {2018})}\BibitemShut {NoStop}%
\bibitem [{\citenamefont {Hertzog}\ \emph {et~al.}(2019)\citenamefont
  {Hertzog}, \citenamefont {Wang}, \citenamefont {Mony},\ and\ \citenamefont
  {B\"orjesson}}]{cavity}%
  \BibitemOpen
  \bibfield  {author} {\bibinfo {author} {\bibfnamefont {M.}~\bibnamefont
  {Hertzog}}, \bibinfo {author} {\bibfnamefont {M.}~\bibnamefont {Wang}},
  \bibinfo {author} {\bibfnamefont {J.}~\bibnamefont {Mony}},\ and\ \bibinfo
  {author} {\bibfnamefont {K.}~\bibnamefont {B\"orjesson}},\ }\bibfield
  {title} {\bibinfo {title} {Strong light-matter interactions: A new direction
  within chemistry},\ }\href@noop {} {\bibfield  {journal} {\bibinfo  {journal}
  {Chem. Soc. Rev.}\ }\textbf {\bibinfo {volume} {48}},\ \bibinfo {pages} {937}
  (\bibinfo {year} {2019})}\BibitemShut {NoStop}%
\bibitem [{\citenamefont {Cassing}\ and\ \citenamefont
  {Bratkovskaya}(2009)}]{high-energy}%
  \BibitemOpen
  \bibfield  {author} {\bibinfo {author} {\bibfnamefont {W.}~\bibnamefont
  {Cassing}}\ and\ \bibinfo {author} {\bibfnamefont {E.}~\bibnamefont
  {Bratkovskaya}},\ }\bibfield  {title} {\bibinfo {title}
  {Parton–hadron–string dynamics: An off-shell transport approach for
  relativistic energies},\ }\href@noop {} {\bibfield  {journal} {\bibinfo
  {journal} {Nucl. Phys. A}\ }\textbf {\bibinfo {volume} {831}},\ \bibinfo
  {pages} {215} (\bibinfo {year} {2009})}\BibitemShut {NoStop}%
\bibitem [{\citenamefont {Haroche}(2013)}]{haroche2013nobel}%
  \BibitemOpen
  \bibfield  {author} {\bibinfo {author} {\bibfnamefont {S.}~\bibnamefont
  {Haroche}},\ }\bibfield  {title} {\bibinfo {title} {Nobel lecture:
  Controlling photons in a box and exploring the quantum to classical
  boundary},\ }\href@noop {} {\bibfield  {journal} {\bibinfo  {journal}
  {Reviews of Modern Physics}\ }\textbf {\bibinfo {volume} {85}},\ \bibinfo
  {pages} {1083} (\bibinfo {year} {2013})}\BibitemShut {NoStop}%
\bibitem [{\citenamefont {Basov}\ \emph {et~al.}(2017)\citenamefont {Basov},
  \citenamefont {Averitt},\ and\ \citenamefont {Hsieh}}]{basov2017towards}%
  \BibitemOpen
  \bibfield  {author} {\bibinfo {author} {\bibfnamefont {D.}~\bibnamefont
  {Basov}}, \bibinfo {author} {\bibfnamefont {R.}~\bibnamefont {Averitt}},\
  and\ \bibinfo {author} {\bibfnamefont {D.}~\bibnamefont {Hsieh}},\ }\bibfield
   {title} {\bibinfo {title} {Towards properties on demand in quantum
  materials},\ }\href@noop {} {\bibfield  {journal} {\bibinfo  {journal}
  {Nature materials}\ }\textbf {\bibinfo {volume} {16}},\ \bibinfo {pages}
  {1077} (\bibinfo {year} {2017})}\BibitemShut {NoStop}%
\bibitem [{\citenamefont {Arute}\ \emph {et~al.}(2020)\citenamefont {Arute},
  \citenamefont {Arya}, \citenamefont {Babbush}, \citenamefont {Bacon},
  \citenamefont {Bardin}, \citenamefont {Barends}, \citenamefont {Bengtsson},
  \citenamefont {Boixo}, \citenamefont {Broughton}, \citenamefont {Buckley},
  \citenamefont {Buell}, \citenamefont {Burkett}, \citenamefont {Bushnell},
  \citenamefont {Chen}, \citenamefont {Chen}, \citenamefont {Chen},
  \citenamefont {Chiaro}, \citenamefont {Collins}, \citenamefont {Cotton},
  \citenamefont {Courtney}, \citenamefont {Demura}, \citenamefont {Derk},
  \citenamefont {Dunsworth}, \citenamefont {Eppens}, \citenamefont {Eckl},
  \citenamefont {Erickson}, \citenamefont {Farhi}, \citenamefont {Fowler},
  \citenamefont {Foxen}, \citenamefont {Gidney}, \citenamefont {Giustina},
  \citenamefont {Graff}, \citenamefont {Gross}, \citenamefont {Habegger},
  \citenamefont {Harrigan}, \citenamefont {Ho}, \citenamefont {Hong},
  \citenamefont {Huang}, \citenamefont {Huggins}, \citenamefont {Ioffe},
  \citenamefont {Isakov}, \citenamefont {Jeffrey}, \citenamefont {Jiang},
  \citenamefont {Jones}, \citenamefont {Kafri}, \citenamefont {Kechedzhi},
  \citenamefont {Kelly}, \citenamefont {Kim}, \citenamefont {Klimov},
  \citenamefont {Korotkov}, \citenamefont {Kostritsa}, \citenamefont
  {Landhuis}, \citenamefont {Laptev}, \citenamefont {Lindmark}, \citenamefont
  {Lucero}, \citenamefont {Marthaler}, \citenamefont {Martin}, \citenamefont
  {Martinis}, \citenamefont {Marusczyk}, \citenamefont {McArdle}, \citenamefont
  {McClean}, \citenamefont {McCourt}, \citenamefont {McEwen}, \citenamefont
  {Megrant}, \citenamefont {Mejuto-Zaera}, \citenamefont {Mi}, \citenamefont
  {Mohseni}, \citenamefont {Mruczkiewicz}, \citenamefont {Mutus}, \citenamefont
  {Naaman}, \citenamefont {Neeley}, \citenamefont {Neill}, \citenamefont
  {Neven}, \citenamefont {Newman}, \citenamefont {Niu}, \citenamefont
  {O'Brien}, \citenamefont {Ostby}, \citenamefont {Pató}, \citenamefont
  {Petukhov}, \citenamefont {Putterman}, \citenamefont {Quintana},
  \citenamefont {Reiner}, \citenamefont {Roushan}, \citenamefont {Rubin},
  \citenamefont {Sank}, \citenamefont {Satzinger}, \citenamefont {Smelyanskiy},
  \citenamefont {Strain}, \citenamefont {Sung}, \citenamefont {Schmitteckert},
  \citenamefont {Szalay}, \citenamefont {Tubman}, \citenamefont {Vainsencher},
  \citenamefont {White}, \citenamefont {Vogt}, \citenamefont {Yao},
  \citenamefont {Yeh}, \citenamefont {Zalcman},\ and\ \citenamefont
  {Zanker}}]{arute2020observation}%
  \BibitemOpen
  \bibfield  {author} {\bibinfo {author} {\bibfnamefont {F.}~\bibnamefont
  {Arute}}, \bibinfo {author} {\bibfnamefont {K.}~\bibnamefont {Arya}},
  \bibinfo {author} {\bibfnamefont {R.}~\bibnamefont {Babbush}}, \bibinfo
  {author} {\bibfnamefont {D.}~\bibnamefont {Bacon}}, \bibinfo {author}
  {\bibfnamefont {J.~C.}\ \bibnamefont {Bardin}}, \bibinfo {author}
  {\bibfnamefont {R.}~\bibnamefont {Barends}}, \bibinfo {author} {\bibfnamefont
  {A.}~\bibnamefont {Bengtsson}}, \bibinfo {author} {\bibfnamefont
  {S.}~\bibnamefont {Boixo}}, \bibinfo {author} {\bibfnamefont
  {M.}~\bibnamefont {Broughton}}, \bibinfo {author} {\bibfnamefont {B.~B.}\
  \bibnamefont {Buckley}}, \bibinfo {author} {\bibfnamefont {D.~A.}\
  \bibnamefont {Buell}}, \bibinfo {author} {\bibfnamefont {B.}~\bibnamefont
  {Burkett}}, \bibinfo {author} {\bibfnamefont {N.}~\bibnamefont {Bushnell}},
  \bibinfo {author} {\bibfnamefont {Y.}~\bibnamefont {Chen}}, \bibinfo {author}
  {\bibfnamefont {Z.}~\bibnamefont {Chen}}, \bibinfo {author} {\bibfnamefont
  {Y.-A.}\ \bibnamefont {Chen}}, \bibinfo {author} {\bibfnamefont
  {B.}~\bibnamefont {Chiaro}}, \bibinfo {author} {\bibfnamefont
  {R.}~\bibnamefont {Collins}}, \bibinfo {author} {\bibfnamefont {S.~J.}\
  \bibnamefont {Cotton}}, \bibinfo {author} {\bibfnamefont {W.}~\bibnamefont
  {Courtney}}, \bibinfo {author} {\bibfnamefont {S.}~\bibnamefont {Demura}},
  \bibinfo {author} {\bibfnamefont {A.}~\bibnamefont {Derk}}, \bibinfo {author}
  {\bibfnamefont {A.}~\bibnamefont {Dunsworth}}, \bibinfo {author}
  {\bibfnamefont {D.}~\bibnamefont {Eppens}}, \bibinfo {author} {\bibfnamefont
  {T.}~\bibnamefont {Eckl}}, \bibinfo {author} {\bibfnamefont {C.}~\bibnamefont
  {Erickson}}, \bibinfo {author} {\bibfnamefont {E.}~\bibnamefont {Farhi}},
  \bibinfo {author} {\bibfnamefont {A.}~\bibnamefont {Fowler}}, \bibinfo
  {author} {\bibfnamefont {B.}~\bibnamefont {Foxen}}, \bibinfo {author}
  {\bibfnamefont {C.}~\bibnamefont {Gidney}}, \bibinfo {author} {\bibfnamefont
  {M.}~\bibnamefont {Giustina}}, \bibinfo {author} {\bibfnamefont
  {R.}~\bibnamefont {Graff}}, \bibinfo {author} {\bibfnamefont {J.~A.}\
  \bibnamefont {Gross}}, \bibinfo {author} {\bibfnamefont {S.}~\bibnamefont
  {Habegger}}, \bibinfo {author} {\bibfnamefont {M.~P.}\ \bibnamefont
  {Harrigan}}, \bibinfo {author} {\bibfnamefont {A.}~\bibnamefont {Ho}},
  \bibinfo {author} {\bibfnamefont {S.}~\bibnamefont {Hong}}, \bibinfo {author}
  {\bibfnamefont {T.}~\bibnamefont {Huang}}, \bibinfo {author} {\bibfnamefont
  {W.}~\bibnamefont {Huggins}}, \bibinfo {author} {\bibfnamefont {L.~B.}\
  \bibnamefont {Ioffe}}, \bibinfo {author} {\bibfnamefont {S.~V.}\ \bibnamefont
  {Isakov}}, \bibinfo {author} {\bibfnamefont {E.}~\bibnamefont {Jeffrey}},
  \bibinfo {author} {\bibfnamefont {Z.}~\bibnamefont {Jiang}}, \bibinfo
  {author} {\bibfnamefont {C.}~\bibnamefont {Jones}}, \bibinfo {author}
  {\bibfnamefont {D.}~\bibnamefont {Kafri}}, \bibinfo {author} {\bibfnamefont
  {K.}~\bibnamefont {Kechedzhi}}, \bibinfo {author} {\bibfnamefont
  {J.}~\bibnamefont {Kelly}}, \bibinfo {author} {\bibfnamefont
  {S.}~\bibnamefont {Kim}}, \bibinfo {author} {\bibfnamefont {P.~V.}\
  \bibnamefont {Klimov}}, \bibinfo {author} {\bibfnamefont {A.~N.}\
  \bibnamefont {Korotkov}}, \bibinfo {author} {\bibfnamefont {F.}~\bibnamefont
  {Kostritsa}}, \bibinfo {author} {\bibfnamefont {D.}~\bibnamefont {Landhuis}},
  \bibinfo {author} {\bibfnamefont {P.}~\bibnamefont {Laptev}}, \bibinfo
  {author} {\bibfnamefont {M.}~\bibnamefont {Lindmark}}, \bibinfo {author}
  {\bibfnamefont {E.}~\bibnamefont {Lucero}}, \bibinfo {author} {\bibfnamefont
  {M.}~\bibnamefont {Marthaler}}, \bibinfo {author} {\bibfnamefont
  {O.}~\bibnamefont {Martin}}, \bibinfo {author} {\bibfnamefont {J.~M.}\
  \bibnamefont {Martinis}}, \bibinfo {author} {\bibfnamefont {A.}~\bibnamefont
  {Marusczyk}}, \bibinfo {author} {\bibfnamefont {S.}~\bibnamefont {McArdle}},
  \bibinfo {author} {\bibfnamefont {J.~R.}\ \bibnamefont {McClean}}, \bibinfo
  {author} {\bibfnamefont {T.}~\bibnamefont {McCourt}}, \bibinfo {author}
  {\bibfnamefont {M.}~\bibnamefont {McEwen}}, \bibinfo {author} {\bibfnamefont
  {A.}~\bibnamefont {Megrant}}, \bibinfo {author} {\bibfnamefont
  {C.}~\bibnamefont {Mejuto-Zaera}}, \bibinfo {author} {\bibfnamefont
  {X.}~\bibnamefont {Mi}}, \bibinfo {author} {\bibfnamefont {M.}~\bibnamefont
  {Mohseni}}, \bibinfo {author} {\bibfnamefont {W.}~\bibnamefont
  {Mruczkiewicz}}, \bibinfo {author} {\bibfnamefont {J.}~\bibnamefont {Mutus}},
  \bibinfo {author} {\bibfnamefont {O.}~\bibnamefont {Naaman}}, \bibinfo
  {author} {\bibfnamefont {M.}~\bibnamefont {Neeley}}, \bibinfo {author}
  {\bibfnamefont {C.}~\bibnamefont {Neill}}, \bibinfo {author} {\bibfnamefont
  {H.}~\bibnamefont {Neven}}, \bibinfo {author} {\bibfnamefont
  {M.}~\bibnamefont {Newman}}, \bibinfo {author} {\bibfnamefont {M.~Y.}\
  \bibnamefont {Niu}}, \bibinfo {author} {\bibfnamefont {T.~E.}\ \bibnamefont
  {O'Brien}}, \bibinfo {author} {\bibfnamefont {E.}~\bibnamefont {Ostby}},
  \bibinfo {author} {\bibfnamefont {B.}~\bibnamefont {Pató}}, \bibinfo
  {author} {\bibfnamefont {A.}~\bibnamefont {Petukhov}}, \bibinfo {author}
  {\bibfnamefont {H.}~\bibnamefont {Putterman}}, \bibinfo {author}
  {\bibfnamefont {C.}~\bibnamefont {Quintana}}, \bibinfo {author}
  {\bibfnamefont {J.-M.}\ \bibnamefont {Reiner}}, \bibinfo {author}
  {\bibfnamefont {P.}~\bibnamefont {Roushan}}, \bibinfo {author} {\bibfnamefont
  {N.~C.}\ \bibnamefont {Rubin}}, \bibinfo {author} {\bibfnamefont
  {D.}~\bibnamefont {Sank}}, \bibinfo {author} {\bibfnamefont {K.~J.}\
  \bibnamefont {Satzinger}}, \bibinfo {author} {\bibfnamefont {V.}~\bibnamefont
  {Smelyanskiy}}, \bibinfo {author} {\bibfnamefont {D.}~\bibnamefont {Strain}},
  \bibinfo {author} {\bibfnamefont {K.~J.}\ \bibnamefont {Sung}}, \bibinfo
  {author} {\bibfnamefont {P.}~\bibnamefont {Schmitteckert}}, \bibinfo {author}
  {\bibfnamefont {M.}~\bibnamefont {Szalay}}, \bibinfo {author} {\bibfnamefont
  {N.~M.}\ \bibnamefont {Tubman}}, \bibinfo {author} {\bibfnamefont
  {A.}~\bibnamefont {Vainsencher}}, \bibinfo {author} {\bibfnamefont
  {T.}~\bibnamefont {White}}, \bibinfo {author} {\bibfnamefont
  {N.}~\bibnamefont {Vogt}}, \bibinfo {author} {\bibfnamefont {Z.~J.}\
  \bibnamefont {Yao}}, \bibinfo {author} {\bibfnamefont {P.}~\bibnamefont
  {Yeh}}, \bibinfo {author} {\bibfnamefont {A.}~\bibnamefont {Zalcman}},\ and\
  \bibinfo {author} {\bibfnamefont {S.}~\bibnamefont {Zanker}},\ }\bibfield
  {title} {\bibinfo {title} {Observation of separated dynamics of charge and
  spin in the fermi-hubbard model},\ }\href@noop {} {\bibfield  {journal}
  {\bibinfo  {journal} {arXiv preprint arXiv:2010.07965}\ } (\bibinfo {year}
  {2020})}\BibitemShut {NoStop}%
\bibitem [{\citenamefont {Del~Re}\ \emph {et~al.}(2020)\citenamefont {Del~Re},
  \citenamefont {Rost}, \citenamefont {Kemper},\ and\ \citenamefont
  {Freericks}}]{del2020driven}%
  \BibitemOpen
  \bibfield  {author} {\bibinfo {author} {\bibfnamefont {L.}~\bibnamefont
  {Del~Re}}, \bibinfo {author} {\bibfnamefont {B.}~\bibnamefont {Rost}},
  \bibinfo {author} {\bibfnamefont {A.}~\bibnamefont {Kemper}},\ and\ \bibinfo
  {author} {\bibfnamefont {J.}~\bibnamefont {Freericks}},\ }\bibfield  {title}
  {\bibinfo {title} {Driven-dissipative quantum mechanics on a lattice:
  Simulating a fermionic reservoir on a quantum computer},\ }\href@noop {}
  {\bibfield  {journal} {\bibinfo  {journal} {Phys. Rev. B}\ }\textbf {\bibinfo
  {volume} {102}},\ \bibinfo {pages} {125112} (\bibinfo {year}
  {2020})}\BibitemShut {NoStop}%
\bibitem [{\citenamefont {Zanardi}\ and\ \citenamefont
  {Venuti}(2015)}]{zanardi2015geometry}%
  \BibitemOpen
  \bibfield  {author} {\bibinfo {author} {\bibfnamefont {P.}~\bibnamefont
  {Zanardi}}\ and\ \bibinfo {author} {\bibfnamefont {L.~C.}\ \bibnamefont
  {Venuti}},\ }\bibfield  {title} {\bibinfo {title} {Geometry, robustness, and
  emerging unitarity in dissipation-projected dynamics},\ }\href@noop {}
  {\bibfield  {journal} {\bibinfo  {journal} {Phys. Rev. A}\ }\textbf {\bibinfo
  {volume} {91}},\ \bibinfo {pages} {052324} (\bibinfo {year}
  {2015})}\BibitemShut {NoStop}%
\bibitem [{\citenamefont {Schirmer}\ and\ \citenamefont
  {Wang}(2010)}]{schirmer2010stabilizing}%
  \BibitemOpen
  \bibfield  {author} {\bibinfo {author} {\bibfnamefont {S.}~\bibnamefont
  {Schirmer}}\ and\ \bibinfo {author} {\bibfnamefont {X.}~\bibnamefont
  {Wang}},\ }\bibfield  {title} {\bibinfo {title} {Stabilizing open quantum
  systems by markovian reservoir engineering},\ }\href@noop {} {\bibfield
  {journal} {\bibinfo  {journal} {Phys. Rev. A}\ }\textbf {\bibinfo {volume}
  {81}},\ \bibinfo {pages} {062306} (\bibinfo {year} {2010})}\BibitemShut
  {NoStop}%
\bibitem [{\citenamefont {Jaschke}\ \emph {et~al.}(2019)\citenamefont
  {Jaschke}, \citenamefont {Carr},\ and\ \citenamefont
  {de~Vega}}]{jaschke2019thermalization}%
  \BibitemOpen
  \bibfield  {author} {\bibinfo {author} {\bibfnamefont {D.}~\bibnamefont
  {Jaschke}}, \bibinfo {author} {\bibfnamefont {L.~D.}\ \bibnamefont {Carr}},\
  and\ \bibinfo {author} {\bibfnamefont {I.}~\bibnamefont {de~Vega}},\
  }\bibfield  {title} {\bibinfo {title} {Thermalization in the quantum ising
  model—approximations, limits, and beyond},\ }\href@noop {} {\bibfield
  {journal} {\bibinfo  {journal} {Quantum Science and Technology}\ }\textbf
  {\bibinfo {volume} {4}},\ \bibinfo {pages} {034002} (\bibinfo {year}
  {2019})}\BibitemShut {NoStop}%
\bibitem [{\citenamefont {Hu}\ \emph {et~al.}(2020{\natexlab{a}})\citenamefont
  {Hu}, \citenamefont {Xia},\ and\ \citenamefont {Kais}}]{hu2020quantum}%
  \BibitemOpen
  \bibfield  {author} {\bibinfo {author} {\bibfnamefont {Z.}~\bibnamefont
  {Hu}}, \bibinfo {author} {\bibfnamefont {R.}~\bibnamefont {Xia}},\ and\
  \bibinfo {author} {\bibfnamefont {S.}~\bibnamefont {Kais}},\ }\bibfield
  {title} {\bibinfo {title} {A quantum algorithm for evolving open quantum
  dynamics on quantum computing devices},\ }\href@noop {} {\bibfield  {journal}
  {\bibinfo  {journal} {Scientific reports}\ }\textbf {\bibinfo {volume}
  {10}},\ \bibinfo {pages} {3301} (\bibinfo {year}
  {2020}{\natexlab{a}})}\BibitemShut {NoStop}%
\bibitem [{\citenamefont {Rost}\ \emph {et~al.}(2020)\citenamefont {Rost},
  \citenamefont {Jones}, \citenamefont {Vyushkova}, \citenamefont {Ali},
  \citenamefont {Cullip}, \citenamefont {Vyushkov},\ and\ \citenamefont
  {Nabrzyski}}]{rost2020simulation}%
  \BibitemOpen
  \bibfield  {author} {\bibinfo {author} {\bibfnamefont {B.}~\bibnamefont
  {Rost}}, \bibinfo {author} {\bibfnamefont {B.}~\bibnamefont {Jones}},
  \bibinfo {author} {\bibfnamefont {M.}~\bibnamefont {Vyushkova}}, \bibinfo
  {author} {\bibfnamefont {A.}~\bibnamefont {Ali}}, \bibinfo {author}
  {\bibfnamefont {C.}~\bibnamefont {Cullip}}, \bibinfo {author} {\bibfnamefont
  {A.}~\bibnamefont {Vyushkov}},\ and\ \bibinfo {author} {\bibfnamefont
  {J.}~\bibnamefont {Nabrzyski}},\ }\bibfield  {title} {\bibinfo {title}
  {Simulation of thermal relaxation in spin chemistry systems on a quantum
  computer using inherent qubit decoherence},\ }\href@noop {} {\bibfield
  {journal} {\bibinfo  {journal} {arXiv preprint arXiv:2001.00794}\ } (\bibinfo
  {year} {2020})}\BibitemShut {NoStop}%
\bibitem [{\citenamefont {Tacchino}\ \emph {et~al.}(2020)\citenamefont
  {Tacchino}, \citenamefont {Chiesa}, \citenamefont {Carretta},\ and\
  \citenamefont {Gerace}}]{tacchino2020quantum}%
  \BibitemOpen
  \bibfield  {author} {\bibinfo {author} {\bibfnamefont {F.}~\bibnamefont
  {Tacchino}}, \bibinfo {author} {\bibfnamefont {A.}~\bibnamefont {Chiesa}},
  \bibinfo {author} {\bibfnamefont {S.}~\bibnamefont {Carretta}},\ and\
  \bibinfo {author} {\bibfnamefont {D.}~\bibnamefont {Gerace}},\ }\bibfield
  {title} {\bibinfo {title} {Quantum computers as universal quantum simulators:
  state-of-the-art and perspectives},\ }\href@noop {} {\bibfield  {journal}
  {\bibinfo  {journal} {Advanced Quantum Technologies}\ }\textbf {\bibinfo
  {volume} {3}},\ \bibinfo {pages} {1900052} (\bibinfo {year}
  {2020})}\BibitemShut {NoStop}%
\bibitem [{\citenamefont {de~Jong}\ \emph {et~al.}(2021)\citenamefont
  {de~Jong}, \citenamefont {Metcalf}, \citenamefont {Mulligan}, \citenamefont
  {P{\l}osko{\'n}}, \citenamefont {Ringer},\ and\ \citenamefont
  {Yao}}]{de2021quantum}%
  \BibitemOpen
  \bibfield  {author} {\bibinfo {author} {\bibfnamefont {W.~A.}\ \bibnamefont
  {de~Jong}}, \bibinfo {author} {\bibfnamefont {M.}~\bibnamefont {Metcalf}},
  \bibinfo {author} {\bibfnamefont {J.}~\bibnamefont {Mulligan}}, \bibinfo
  {author} {\bibfnamefont {M.}~\bibnamefont {P{\l}osko{\'n}}}, \bibinfo
  {author} {\bibfnamefont {F.}~\bibnamefont {Ringer}},\ and\ \bibinfo {author}
  {\bibfnamefont {X.}~\bibnamefont {Yao}},\ }\bibfield  {title} {\bibinfo
  {title} {Quantum simulation of open quantum systems in heavy-ion
  collisions},\ }\href@noop {} {\bibfield  {journal} {\bibinfo  {journal}
  {Physical Review D}\ }\textbf {\bibinfo {volume} {104}},\ \bibinfo {pages}
  {L051501} (\bibinfo {year} {2021})}\BibitemShut {NoStop}%
\bibitem [{\citenamefont {Ramusat}\ and\ \citenamefont
  {Savona}(2021)}]{ramusat2021quantum}%
  \BibitemOpen
  \bibfield  {author} {\bibinfo {author} {\bibfnamefont {N.}~\bibnamefont
  {Ramusat}}\ and\ \bibinfo {author} {\bibfnamefont {V.}~\bibnamefont
  {Savona}},\ }\bibfield  {title} {\bibinfo {title} {A quantum algorithm for
  the direct estimation of the steady state of open quantum systems},\
  }\href@noop {} {\bibfield  {journal} {\bibinfo  {journal} {Quantum}\ }\textbf
  {\bibinfo {volume} {5}},\ \bibinfo {pages} {399} (\bibinfo {year}
  {2021})}\BibitemShut {NoStop}%
\bibitem [{\citenamefont {Schlimgen}\ \emph {et~al.}(2022)\citenamefont
  {Schlimgen}, \citenamefont {Head-Marsden}, \citenamefont {Sager},
  \citenamefont {Narang},\ and\ \citenamefont
  {Mazziotti}}]{schlimgen2022quantum}%
  \BibitemOpen
  \bibfield  {author} {\bibinfo {author} {\bibfnamefont {A.~W.}\ \bibnamefont
  {Schlimgen}}, \bibinfo {author} {\bibfnamefont {K.}~\bibnamefont
  {Head-Marsden}}, \bibinfo {author} {\bibfnamefont {L.~M.}\ \bibnamefont
  {Sager}}, \bibinfo {author} {\bibfnamefont {P.}~\bibnamefont {Narang}},\ and\
  \bibinfo {author} {\bibfnamefont {D.~A.}\ \bibnamefont {Mazziotti}},\
  }\bibfield  {title} {\bibinfo {title} {Quantum simulation of the lindblad
  equation using a unitary decomposition of operators},\ }\href@noop {}
  {\bibfield  {journal} {\bibinfo  {journal} {Physical Review Research}\
  }\textbf {\bibinfo {volume} {4}},\ \bibinfo {pages} {023216} (\bibinfo {year}
  {2022})}\BibitemShut {NoStop}%
\bibitem [{\citenamefont {Hu}\ \emph {et~al.}(2022{\natexlab{a}})\citenamefont
  {Hu}, \citenamefont {Head-Marsden}, \citenamefont {Mazziotti}, \citenamefont
  {Narang},\ and\ \citenamefont {Kais}}]{hu2022general}%
  \BibitemOpen
  \bibfield  {author} {\bibinfo {author} {\bibfnamefont {Z.}~\bibnamefont
  {Hu}}, \bibinfo {author} {\bibfnamefont {K.}~\bibnamefont {Head-Marsden}},
  \bibinfo {author} {\bibfnamefont {D.~A.}\ \bibnamefont {Mazziotti}}, \bibinfo
  {author} {\bibfnamefont {P.}~\bibnamefont {Narang}},\ and\ \bibinfo {author}
  {\bibfnamefont {S.}~\bibnamefont {Kais}},\ }\bibfield  {title} {\bibinfo
  {title} {A general quantum algorithm for open quantum dynamics demonstrated
  with the fenna-matthews-olson complex},\ }\href@noop {} {\bibfield  {journal}
  {\bibinfo  {journal} {Quantum}\ }\textbf {\bibinfo {volume} {6}},\ \bibinfo
  {pages} {726} (\bibinfo {year} {2022}{\natexlab{a}})}\BibitemShut {NoStop}%
\bibitem [{\citenamefont {Wang}\ \emph {et~al.}(2023)\citenamefont {Wang},
  \citenamefont {Mulvihill}, \citenamefont {Hu}, \citenamefont {Lyu},
  \citenamefont {Shivpuje}, \citenamefont {Liu}, \citenamefont {Soley},
  \citenamefont {Geva}, \citenamefont {Batista},\ and\ \citenamefont
  {Kais}}]{wang2023simulating}%
  \BibitemOpen
  \bibfield  {author} {\bibinfo {author} {\bibfnamefont {Y.}~\bibnamefont
  {Wang}}, \bibinfo {author} {\bibfnamefont {E.}~\bibnamefont {Mulvihill}},
  \bibinfo {author} {\bibfnamefont {Z.}~\bibnamefont {Hu}}, \bibinfo {author}
  {\bibfnamefont {N.}~\bibnamefont {Lyu}}, \bibinfo {author} {\bibfnamefont
  {S.}~\bibnamefont {Shivpuje}}, \bibinfo {author} {\bibfnamefont
  {Y.}~\bibnamefont {Liu}}, \bibinfo {author} {\bibfnamefont {M.~B.}\
  \bibnamefont {Soley}}, \bibinfo {author} {\bibfnamefont {E.}~\bibnamefont
  {Geva}}, \bibinfo {author} {\bibfnamefont {V.~S.}\ \bibnamefont {Batista}},\
  and\ \bibinfo {author} {\bibfnamefont {S.}~\bibnamefont {Kais}},\ }\bibfield
  {title} {\bibinfo {title} {Simulating open quantum system dynamics on nisq
  computers with generalized quantum master equations},\ }\href@noop {}
  {\bibfield  {journal} {\bibinfo  {journal} {Journal of Chemical Theory and
  Computation}\ } (\bibinfo {year} {2023})}\BibitemShut {NoStop}%
\bibitem [{\citenamefont {Mi}\ \emph {et~al.}(2023)\citenamefont {Mi},
  \citenamefont {Michailidis}, \citenamefont {Shabani}, \citenamefont {Miao},
  \citenamefont {Klimov}, \citenamefont {Lloyd}, \citenamefont {Rosenberg},
  \citenamefont {Acharya}, \citenamefont {Aleiner}, \citenamefont {Andersen}
  \emph {et~al.}}]{mi2023stable}%
  \BibitemOpen
  \bibfield  {author} {\bibinfo {author} {\bibfnamefont {X.}~\bibnamefont
  {Mi}}, \bibinfo {author} {\bibfnamefont {A.}~\bibnamefont {Michailidis}},
  \bibinfo {author} {\bibfnamefont {S.}~\bibnamefont {Shabani}}, \bibinfo
  {author} {\bibfnamefont {K.}~\bibnamefont {Miao}}, \bibinfo {author}
  {\bibfnamefont {P.}~\bibnamefont {Klimov}}, \bibinfo {author} {\bibfnamefont
  {J.}~\bibnamefont {Lloyd}}, \bibinfo {author} {\bibfnamefont
  {E.}~\bibnamefont {Rosenberg}}, \bibinfo {author} {\bibfnamefont
  {R.}~\bibnamefont {Acharya}}, \bibinfo {author} {\bibfnamefont
  {I.}~\bibnamefont {Aleiner}}, \bibinfo {author} {\bibfnamefont
  {T.}~\bibnamefont {Andersen}}, \emph {et~al.},\ }\bibfield  {title} {\bibinfo
  {title} {Stable quantum-correlated many body states via engineered
  dissipation},\ }\href@noop {} {\bibfield  {journal} {\bibinfo  {journal}
  {arXiv preprint arXiv:2304.13878}\ } (\bibinfo {year} {2023})}\BibitemShut
  {NoStop}%
\bibitem [{\citenamefont {Breuer}\ and\ \citenamefont
  {Petruccione}(2002)}]{breuer2002theory}%
  \BibitemOpen
  \bibfield  {author} {\bibinfo {author} {\bibfnamefont {H.-P.}\ \bibnamefont
  {Breuer}}\ and\ \bibinfo {author} {\bibfnamefont {F.}~\bibnamefont
  {Petruccione}},\ }\href@noop {} {\emph {\bibinfo {title} {The theory of open
  quantum systems}}}\ (\bibinfo  {publisher} {Oxford University Press on
  Demand},\ \bibinfo {address} {New York, NY},\ \bibinfo {year}
  {2002})\BibitemShut {NoStop}%
\bibitem [{\citenamefont {Sommer}\ \emph {et~al.}(2021)\citenamefont {Sommer},
  \citenamefont {Piazza},\ and\ \citenamefont {Luitz}}]{sommer2021}%
  \BibitemOpen
  \bibfield  {author} {\bibinfo {author} {\bibfnamefont {O.~E.}\ \bibnamefont
  {Sommer}}, \bibinfo {author} {\bibfnamefont {F.}~\bibnamefont {Piazza}},\
  and\ \bibinfo {author} {\bibfnamefont {D.~J.}\ \bibnamefont {Luitz}},\
  }\bibfield  {title} {\bibinfo {title} {Many-body hierarchy of dissipative
  timescales in a quantum computer},\ }\href
  {https://doi.org/10.1103/PhysRevResearch.3.023190} {\bibfield  {journal}
  {\bibinfo  {journal} {Phys. Rev. Research}\ }\textbf {\bibinfo {volume}
  {3}},\ \bibinfo {pages} {023190} (\bibinfo {year} {2021})}\BibitemShut
  {NoStop}%
\bibitem [{\citenamefont {Tseng}\ \emph {et~al.}(2000)\citenamefont {Tseng},
  \citenamefont {Somaroo}, \citenamefont {Sharf}, \citenamefont {Knill},
  \citenamefont {Laflamme}, \citenamefont {Havel},\ and\ \citenamefont
  {Cory}}]{tseng2000quantum}%
  \BibitemOpen
  \bibfield  {author} {\bibinfo {author} {\bibfnamefont {C.}~\bibnamefont
  {Tseng}}, \bibinfo {author} {\bibfnamefont {S.}~\bibnamefont {Somaroo}},
  \bibinfo {author} {\bibfnamefont {Y.}~\bibnamefont {Sharf}}, \bibinfo
  {author} {\bibfnamefont {E.}~\bibnamefont {Knill}}, \bibinfo {author}
  {\bibfnamefont {R.}~\bibnamefont {Laflamme}}, \bibinfo {author}
  {\bibfnamefont {T.~F.}\ \bibnamefont {Havel}},\ and\ \bibinfo {author}
  {\bibfnamefont {D.~G.}\ \bibnamefont {Cory}},\ }\bibfield  {title} {\bibinfo
  {title} {Quantum simulation with natural decoherence},\ }\href@noop {}
  {\bibfield  {journal} {\bibinfo  {journal} {Phys. Rev. A}\ }\textbf {\bibinfo
  {volume} {62}},\ \bibinfo {pages} {032309} (\bibinfo {year}
  {2000})}\BibitemShut {NoStop}%
\bibitem [{\citenamefont {Terhal}\ and\ \citenamefont
  {DiVincenzo}(2000)}]{terhal2000problem}%
  \BibitemOpen
  \bibfield  {author} {\bibinfo {author} {\bibfnamefont {B.~M.}\ \bibnamefont
  {Terhal}}\ and\ \bibinfo {author} {\bibfnamefont {D.~P.}\ \bibnamefont
  {DiVincenzo}},\ }\bibfield  {title} {\bibinfo {title} {Problem of
  equilibration and the computation of correlation functions on a quantum
  computer},\ }\href@noop {} {\bibfield  {journal} {\bibinfo  {journal} {Phys.
  Rev. A}\ }\textbf {\bibinfo {volume} {61}},\ \bibinfo {pages} {022301}
  (\bibinfo {year} {2000})}\BibitemShut {NoStop}%
\bibitem [{\citenamefont {Wang}\ \emph {et~al.}(2011)\citenamefont {Wang},
  \citenamefont {Ashhab},\ and\ \citenamefont {Nori}}]{wang2011quantum}%
  \BibitemOpen
  \bibfield  {author} {\bibinfo {author} {\bibfnamefont {H.}~\bibnamefont
  {Wang}}, \bibinfo {author} {\bibfnamefont {S.}~\bibnamefont {Ashhab}},\ and\
  \bibinfo {author} {\bibfnamefont {F.}~\bibnamefont {Nori}},\ }\bibfield
  {title} {\bibinfo {title} {Quantum algorithm for simulating the dynamics of
  an open quantum system},\ }\href@noop {} {\bibfield  {journal} {\bibinfo
  {journal} {Phys. Rev. A}\ }\textbf {\bibinfo {volume} {83}},\ \bibinfo
  {pages} {062317} (\bibinfo {year} {2011})}\BibitemShut {NoStop}%
\bibitem [{\citenamefont {Su}\ and\ \citenamefont {Li}(2020)}]{su2020quantum}%
  \BibitemOpen
  \bibfield  {author} {\bibinfo {author} {\bibfnamefont {H.-Y.}\ \bibnamefont
  {Su}}\ and\ \bibinfo {author} {\bibfnamefont {Y.}~\bibnamefont {Li}},\
  }\bibfield  {title} {\bibinfo {title} {Quantum algorithm for the simulation
  of open-system dynamics and thermalization},\ }\href@noop {} {\bibfield
  {journal} {\bibinfo  {journal} {Phys. Rev. A}\ }\textbf {\bibinfo {volume}
  {101}},\ \bibinfo {pages} {012328} (\bibinfo {year} {2020})}\BibitemShut
  {NoStop}%
\bibitem [{\citenamefont {Cleve}\ and\ \citenamefont
  {Wang}(2016)}]{cleve2016efficient}%
  \BibitemOpen
  \bibfield  {author} {\bibinfo {author} {\bibfnamefont {R.}~\bibnamefont
  {Cleve}}\ and\ \bibinfo {author} {\bibfnamefont {C.}~\bibnamefont {Wang}},\
  }\bibfield  {title} {\bibinfo {title} {Efficient quantum algorithms for
  simulating lindblad evolution},\ }\href@noop {} {\bibfield  {journal}
  {\bibinfo  {journal} {arXiv preprint arXiv:1612.09512}\ } (\bibinfo {year}
  {2016})}\BibitemShut {NoStop}%
\bibitem [{\citenamefont {Hu}\ \emph {et~al.}(2020{\natexlab{b}})\citenamefont
  {Hu}, \citenamefont {Xia},\ and\ \citenamefont {Kais}}]{hu2020}%
  \BibitemOpen
  \bibfield  {author} {\bibinfo {author} {\bibfnamefont {Z.}~\bibnamefont
  {Hu}}, \bibinfo {author} {\bibfnamefont {R.}~\bibnamefont {Xia}},\ and\
  \bibinfo {author} {\bibfnamefont {S.}~\bibnamefont {Kais}},\ }\bibfield
  {title} {\bibinfo {title} {A quantum algorithm for evolving open quantum
  dynamics on quantum computing devices},\ }\href
  {https://doi.org/10.1038/s41598-020-60321-x} {\bibfield  {journal} {\bibinfo
  {journal} {Scientific reports}\ }\textbf {\bibinfo {volume} {10}},\ \bibinfo
  {pages} {1} (\bibinfo {year} {2020}{\natexlab{b}})}\BibitemShut {NoStop}%
\bibitem [{\citenamefont {Hu}\ \emph {et~al.}(2022{\natexlab{b}})\citenamefont
  {Hu}, \citenamefont {Head-Marsden}, \citenamefont {Mazziotti}, \citenamefont
  {Narang},\ and\ \citenamefont {Kais}}]{hu2022}%
  \BibitemOpen
  \bibfield  {author} {\bibinfo {author} {\bibfnamefont {Z.}~\bibnamefont
  {Hu}}, \bibinfo {author} {\bibfnamefont {K.}~\bibnamefont {Head-Marsden}},
  \bibinfo {author} {\bibfnamefont {D.~A.}\ \bibnamefont {Mazziotti}}, \bibinfo
  {author} {\bibfnamefont {P.}~\bibnamefont {Narang}},\ and\ \bibinfo {author}
  {\bibfnamefont {S.}~\bibnamefont {Kais}},\ }\bibfield  {title} {\bibinfo
  {title} {A general quantum algorithm for open quantum dynamics demonstrated
  with the fenna-matthews-olson complex},\ }\href@noop {} {\bibfield  {journal}
  {\bibinfo  {journal} {Quantum}\ }\textbf {\bibinfo {volume} {6}},\ \bibinfo
  {pages} {726} (\bibinfo {year} {2022}{\natexlab{b}})}\BibitemShut {NoStop}%
\bibitem [{\citenamefont {Head-Marsden}\ \emph {et~al.}(2021)\citenamefont
  {Head-Marsden}, \citenamefont {Krastanov}, \citenamefont {Mazziotti},\ and\
  \citenamefont {Narang}}]{head2021capturing}%
  \BibitemOpen
  \bibfield  {author} {\bibinfo {author} {\bibfnamefont {K.}~\bibnamefont
  {Head-Marsden}}, \bibinfo {author} {\bibfnamefont {S.}~\bibnamefont
  {Krastanov}}, \bibinfo {author} {\bibfnamefont {D.~A.}\ \bibnamefont
  {Mazziotti}},\ and\ \bibinfo {author} {\bibfnamefont {P.}~\bibnamefont
  {Narang}},\ }\bibfield  {title} {\bibinfo {title} {Capturing non-markovian
  dynamics on near-term quantum computers},\ }\href@noop {} {\bibfield
  {journal} {\bibinfo  {journal} {Phys. Rev. Research}\ }\textbf {\bibinfo
  {volume} {3}},\ \bibinfo {pages} {013182} (\bibinfo {year}
  {2021})}\BibitemShut {NoStop}%
\bibitem [{\citenamefont {Childs}\ and\ \citenamefont
  {Li}(2017)}]{childs2016efficient}%
  \BibitemOpen
  \bibfield  {author} {\bibinfo {author} {\bibfnamefont {A.~M.}\ \bibnamefont
  {Childs}}\ and\ \bibinfo {author} {\bibfnamefont {T.}~\bibnamefont {Li}},\
  }\bibfield  {title} {\bibinfo {title} {{Efficient simulation of sparse
  Markovian quantum dynamics}},\ }\href {https://doi.org/10.26421/QIC17.11-12}
  {\bibfield  {journal} {\bibinfo  {journal} {Quantum Information and
  Computation}\ }\textbf {\bibinfo {volume} {17}},\ \bibinfo {pages} {901}
  (\bibinfo {year} {2017})},\ \Eprint {https://arxiv.org/abs/1611.05543}
  {arXiv:1611.05543} \BibitemShut {NoStop}%
\bibitem [{\citenamefont {Kliesch}\ \emph {et~al.}(2011)\citenamefont
  {Kliesch}, \citenamefont {Barthel}, \citenamefont {Gogolin}, \citenamefont
  {Kastoryano},\ and\ \citenamefont {Eisert}}]{kliesch2011dissipative}%
  \BibitemOpen
  \bibfield  {author} {\bibinfo {author} {\bibfnamefont {M.}~\bibnamefont
  {Kliesch}}, \bibinfo {author} {\bibfnamefont {T.}~\bibnamefont {Barthel}},
  \bibinfo {author} {\bibfnamefont {C.}~\bibnamefont {Gogolin}}, \bibinfo
  {author} {\bibfnamefont {M.}~\bibnamefont {Kastoryano}},\ and\ \bibinfo
  {author} {\bibfnamefont {J.}~\bibnamefont {Eisert}},\ }\bibfield  {title}
  {\bibinfo {title} {Dissipative quantum church-turing theorem},\ }\href@noop
  {} {\bibfield  {journal} {\bibinfo  {journal} {Phys. Rev. Lett.}\ }\textbf
  {\bibinfo {volume} {107}},\ \bibinfo {pages} {120501} (\bibinfo {year}
  {2011})}\BibitemShut {NoStop}%
\bibitem [{\citenamefont {Haug}\ and\ \citenamefont
  {Bharti}(2020)}]{haug2020generalized}%
  \BibitemOpen
  \bibfield  {author} {\bibinfo {author} {\bibfnamefont {T.}~\bibnamefont
  {Haug}}\ and\ \bibinfo {author} {\bibfnamefont {K.}~\bibnamefont {Bharti}},\
  }\bibfield  {title} {\bibinfo {title} {Generalized quantum assisted
  simulator},\ }\href@noop {} {\bibfield  {journal} {\bibinfo  {journal} {arXiv
  preprint arXiv:2011.14737}\ } (\bibinfo {year} {2020})}\BibitemShut {NoStop}%
\bibitem [{\citenamefont {Yoshioka}\ \emph {et~al.}(2020)\citenamefont
  {Yoshioka}, \citenamefont {Nakagawa}, \citenamefont {Mitarai},\ and\
  \citenamefont {Fujii}}]{yoshioka2020variational}%
  \BibitemOpen
  \bibfield  {author} {\bibinfo {author} {\bibfnamefont {N.}~\bibnamefont
  {Yoshioka}}, \bibinfo {author} {\bibfnamefont {Y.~O.}\ \bibnamefont
  {Nakagawa}}, \bibinfo {author} {\bibfnamefont {K.}~\bibnamefont {Mitarai}},\
  and\ \bibinfo {author} {\bibfnamefont {K.}~\bibnamefont {Fujii}},\ }\bibfield
   {title} {\bibinfo {title} {Variational quantum algorithm for nonequilibrium
  steady states},\ }\href@noop {} {\bibfield  {journal} {\bibinfo  {journal}
  {Phys. Rev. Research}\ }\textbf {\bibinfo {volume} {2}},\ \bibinfo {pages}
  {043289} (\bibinfo {year} {2020})}\BibitemShut {NoStop}%
\bibitem [{\citenamefont {Kamakari}\ \emph {et~al.}(2021)\citenamefont
  {Kamakari}, \citenamefont {Sun}, \citenamefont {Motta},\ and\ \citenamefont
  {Minnich}}]{kamakari2021}%
  \BibitemOpen
  \bibfield  {author} {\bibinfo {author} {\bibfnamefont {H.}~\bibnamefont
  {Kamakari}}, \bibinfo {author} {\bibfnamefont {S.-N.}\ \bibnamefont {Sun}},
  \bibinfo {author} {\bibfnamefont {M.}~\bibnamefont {Motta}},\ and\ \bibinfo
  {author} {\bibfnamefont {A.~J.}\ \bibnamefont {Minnich}},\ }\bibfield
  {title} {\bibinfo {title} {Digital quantum simulation of open quantum systems
  using quantum imaginary time evolution},\ }\href@noop {} {\bibfield
  {journal} {\bibinfo  {journal} {arXiv preprint arXiv:2104.07823}\ } (\bibinfo
  {year} {2021})}\BibitemShut {NoStop}%
\bibitem [{\citenamefont {Metcalf}\ \emph {et~al.}(2020)\citenamefont
  {Metcalf}, \citenamefont {Moussa}, \citenamefont {de~Jong},\ and\
  \citenamefont {Sarovar}}]{metcalf2020engineered}%
  \BibitemOpen
  \bibfield  {author} {\bibinfo {author} {\bibfnamefont {M.}~\bibnamefont
  {Metcalf}}, \bibinfo {author} {\bibfnamefont {J.~E.}\ \bibnamefont {Moussa}},
  \bibinfo {author} {\bibfnamefont {W.~A.}\ \bibnamefont {de~Jong}},\ and\
  \bibinfo {author} {\bibfnamefont {M.}~\bibnamefont {Sarovar}},\ }\bibfield
  {title} {\bibinfo {title} {Engineered thermalization and cooling of quantum
  many-body systems},\ }\href@noop {} {\bibfield  {journal} {\bibinfo
  {journal} {Phys. Rev. Research}\ }\textbf {\bibinfo {volume} {2}},\ \bibinfo
  {pages} {023214} (\bibinfo {year} {2020})}\BibitemShut {NoStop}%
\bibitem [{\citenamefont {Barreiro}\ \emph
  {et~al.}(2011{\natexlab{a}})\citenamefont {Barreiro}, \citenamefont
  {Müller}, \citenamefont {Schindler}, \citenamefont {Nigg}, \citenamefont
  {Monz}, \citenamefont {Chwalla}, \citenamefont {Hennrich}, \citenamefont
  {Roos}, \citenamefont {Zoller},\ and\ \citenamefont {Blatt}}]{Barreiro_2011}%
  \BibitemOpen
  \bibfield  {author} {\bibinfo {author} {\bibfnamefont {J.~T.}\ \bibnamefont
  {Barreiro}}, \bibinfo {author} {\bibfnamefont {M.}~\bibnamefont {Müller}},
  \bibinfo {author} {\bibfnamefont {P.}~\bibnamefont {Schindler}}, \bibinfo
  {author} {\bibfnamefont {D.}~\bibnamefont {Nigg}}, \bibinfo {author}
  {\bibfnamefont {T.}~\bibnamefont {Monz}}, \bibinfo {author} {\bibfnamefont
  {M.}~\bibnamefont {Chwalla}}, \bibinfo {author} {\bibfnamefont
  {M.}~\bibnamefont {Hennrich}}, \bibinfo {author} {\bibfnamefont {C.~F.}\
  \bibnamefont {Roos}}, \bibinfo {author} {\bibfnamefont {P.}~\bibnamefont
  {Zoller}},\ and\ \bibinfo {author} {\bibfnamefont {R.}~\bibnamefont
  {Blatt}},\ }\bibfield  {title} {\bibinfo {title} {An open-system quantum
  simulator with trapped ions},\ }\href {https://doi.org/10.1038/nature09801}
  {\bibfield  {journal} {\bibinfo  {journal} {Nature}\ }\textbf {\bibinfo
  {volume} {470}},\ \bibinfo {pages} {486–491} (\bibinfo {year}
  {2011}{\natexlab{a}})}\BibitemShut {NoStop}%
\bibitem [{\citenamefont {Tornow}\ \emph {et~al.}(2022)\citenamefont {Tornow},
  \citenamefont {Gehrke},\ and\ \citenamefont {Helmbrecht}}]{tornow2022non}%
  \BibitemOpen
  \bibfield  {author} {\bibinfo {author} {\bibfnamefont {S.}~\bibnamefont
  {Tornow}}, \bibinfo {author} {\bibfnamefont {W.}~\bibnamefont {Gehrke}},\
  and\ \bibinfo {author} {\bibfnamefont {U.}~\bibnamefont {Helmbrecht}},\
  }\bibfield  {title} {\bibinfo {title} {Non-equilibrium dynamics of a
  dissipative two-site hubbard model simulated on ibm quantum computers},\
  }\href@noop {} {\bibfield  {journal} {\bibinfo  {journal} {Journal of Physics
  A: Mathematical and Theoretical}\ }\textbf {\bibinfo {volume} {55}},\
  \bibinfo {pages} {245302} (\bibinfo {year} {2022})}\BibitemShut {NoStop}%
\bibitem [{\citenamefont {Han}(2013)}]{han2013solution}%
  \BibitemOpen
  \bibfield  {author} {\bibinfo {author} {\bibfnamefont {J.~E.}\ \bibnamefont
  {Han}},\ }\bibfield  {title} {\bibinfo {title} {Solution of
  electric-field-driven tight-binding lattice coupled to fermion reservoirs},\
  }\href@noop {} {\bibfield  {journal} {\bibinfo  {journal} {Physical Review
  B}\ }\textbf {\bibinfo {volume} {87}},\ \bibinfo {pages} {085119} (\bibinfo
  {year} {2013})}\BibitemShut {NoStop}%
\bibitem [{\citenamefont {Barreiro}\ \emph
  {et~al.}(2011{\natexlab{b}})\citenamefont {Barreiro}, \citenamefont
  {M{\"u}ller}, \citenamefont {Schindler}, \citenamefont {Nigg}, \citenamefont
  {Monz}, \citenamefont {Chwalla}, \citenamefont {Hennrich}, \citenamefont
  {Roos}, \citenamefont {Zoller},\ and\ \citenamefont
  {Blatt}}]{barreiro2011open}%
  \BibitemOpen
  \bibfield  {author} {\bibinfo {author} {\bibfnamefont {J.~T.}\ \bibnamefont
  {Barreiro}}, \bibinfo {author} {\bibfnamefont {M.}~\bibnamefont
  {M{\"u}ller}}, \bibinfo {author} {\bibfnamefont {P.}~\bibnamefont
  {Schindler}}, \bibinfo {author} {\bibfnamefont {D.}~\bibnamefont {Nigg}},
  \bibinfo {author} {\bibfnamefont {T.}~\bibnamefont {Monz}}, \bibinfo {author}
  {\bibfnamefont {M.}~\bibnamefont {Chwalla}}, \bibinfo {author} {\bibfnamefont
  {M.}~\bibnamefont {Hennrich}}, \bibinfo {author} {\bibfnamefont {C.~F.}\
  \bibnamefont {Roos}}, \bibinfo {author} {\bibfnamefont {P.}~\bibnamefont
  {Zoller}},\ and\ \bibinfo {author} {\bibfnamefont {R.}~\bibnamefont
  {Blatt}},\ }\bibfield  {title} {\bibinfo {title} {An open-system quantum
  simulator with trapped ions},\ }\href@noop {} {\bibfield  {journal} {\bibinfo
   {journal} {Nature}\ }\textbf {\bibinfo {volume} {470}},\ \bibinfo {pages}
  {486} (\bibinfo {year} {2011}{\natexlab{b}})}\BibitemShut {NoStop}%
\bibitem [{\citenamefont {Javadi-Abhari}\ \emph {et~al.}(2024)\citenamefont
  {Javadi-Abhari}, \citenamefont {Treinish}, \citenamefont {Krsulich},
  \citenamefont {Wood}, \citenamefont {Lishman}, \citenamefont {Gacon},
  \citenamefont {Martiel}, \citenamefont {Nation}, \citenamefont {Bishop},
  \citenamefont {Cross}, \citenamefont {Johnson},\ and\ \citenamefont
  {Gambetta}}]{qiskit2024}%
  \BibitemOpen
  \bibfield  {author} {\bibinfo {author} {\bibfnamefont {A.}~\bibnamefont
  {Javadi-Abhari}}, \bibinfo {author} {\bibfnamefont {M.}~\bibnamefont
  {Treinish}}, \bibinfo {author} {\bibfnamefont {K.}~\bibnamefont {Krsulich}},
  \bibinfo {author} {\bibfnamefont {C.~J.}\ \bibnamefont {Wood}}, \bibinfo
  {author} {\bibfnamefont {J.}~\bibnamefont {Lishman}}, \bibinfo {author}
  {\bibfnamefont {J.}~\bibnamefont {Gacon}}, \bibinfo {author} {\bibfnamefont
  {S.}~\bibnamefont {Martiel}}, \bibinfo {author} {\bibfnamefont {P.~D.}\
  \bibnamefont {Nation}}, \bibinfo {author} {\bibfnamefont {L.~S.}\
  \bibnamefont {Bishop}}, \bibinfo {author} {\bibfnamefont {A.~W.}\
  \bibnamefont {Cross}}, \bibinfo {author} {\bibfnamefont {B.~R.}\ \bibnamefont
  {Johnson}},\ and\ \bibinfo {author} {\bibfnamefont {J.~M.}\ \bibnamefont
  {Gambetta}},\ }\href {https://doi.org/10.48550/arXiv.2405.08810} {\bibinfo
  {title} {Quantum computing with {Q}iskit}} (\bibinfo {year} {2024}),\ \Eprint
  {https://arxiv.org/abs/2405.08810} {arXiv:2405.08810 [quant-ph]} \BibitemShut
  {NoStop}%
\end{thebibliography}%

\clearpage

\appendix

\renewcommand\theequation{S\arabic{equation}}
\renewcommand\thefigure{S\arabic{figure}}  
\renewcommand\thetable{S\arabic{table}}  
\setcounter{figure}{0}

\section{Technical details of the algorithms}

\subsection{Lattice electrons driven by a DC field}\label{sup:ele}

The open quantum system examined in the first part of the paper is given by non-interacting  fermions on a one-dimensional lattice with nearest-neighbor hopping and driven out of equilibrium by an electric field. The effect of the field is taken into account by introducing a complex Peierls phase  $\varphi(t) = \Omega\,t$ (given by $\Omega = eEa$) to the hopping amplitude $\gamma_h$. The system Hamiltonian then reads:
\begin{equation}\label{ham:1}
\hat{\mathcal H} = -\gamma_h\sum_{i}e^{i\varphi(t)}d^{\dag}_{i}d^{\phantom{\dagger}}_{i+1} + \mbox{h.~c.}
\end{equation}
We let the system interact with a thermostat by coupling every lattice site to an independent fermionic bath, whose Hamiltonian is $\hat{\mathcal H}_b = \sum_{i\alpha}\omega_\alpha c^{\dag}_{i\alpha}c^{\,}_{i\alpha}$, through a  hopping term that is given by:
\begin{equation}\label{ham:2}
\hat{\mathcal V} = -g\sum_{i\alpha} \left(d^\dag_i c^{\,}_{i\alpha} + c^{\dag}_{i\alpha}d^{\,}_i\right).
\end{equation} 
Here $g$ is the bare hybridization amplitude, and $\alpha$ is an index that runs over all the internal degrees of freedom of the bath, which are taken to be infinite.
The total Hamiltonian of the system plus the bath is given by $\hat{\mathcal H}_{tot} = \hat{\mathcal H} +\hat{\mathcal H}_{b} +\hat{\mathcal V} $ and can be block diagonalized by expanding the fields in their Fourier components; that is, $d_k = \frac{1}{\sqrt{N}}\sum_n d_n e^{-ikn}$, $c_{k\alpha} = \frac{1}{\sqrt{N}}\sum_n c_{n\alpha} e^{-ikn}$. In this basis, the Hamiltonian decomposes into a sum of Hamiltonians for each momentum, $\hat{\mathcal{H}}_{tot} =  \sum_k\hat{\mathcal{H}}_{tot}^{(k)} $, with $\hat{\mathcal{H}}_{tot}^{(k)} = \hat{\mathcal{H}}^{(k)}+\hat{\mathcal{H}}_b^{(k)} + \hat{\mathcal{V}}^{(k)}$ and where

\begin{eqnarray}
\label{ham:sys1}
\hat{\mathcal H}^{(k)}(t) &=& -2\gamma_h \cos\left(k+\Omega\,t\right)d^\dag_k d^{\phantom{\dagger}}_{k},\\
\label{ham:bath}
\hat{\mathcal H}_b^{(k)} &=& \sum_\alpha \omega_\alpha c^\dag_{k\alpha}c^{\,}_{k\alpha},\\
\label{ham:hyb1}
\hat{\mathcal V}^{(k)} &=& -g\sum_{\alpha}d^\dag_k c^{\,}_{k\alpha} + \mbox{h.~c.}
\end{eqnarray}
The Lindblad master equation of the density matrix for a given $k$-point then becomes
\begin{equation}\label{eq:me_lin}
    \partial_t \rho_k = -i[\mathcal{H}^{(k)}(t),\rho_k]+\sum_{\ell=\{1,2\}}L_{\ell}^{\,} \,\rho_k \,L_\ell^{\dag} - \{\rho_k,L^\dag_\ell L^{\,}_\ell\},
\end{equation}
where the Lindblad operators are $L_1 = \sqrt{\Gamma\,n_F(\epsilon_k(t))}d_k$ and $L_2 = \sqrt{\Gamma\,n_F(-\epsilon_k(t))}d^\dag_k$, with $\Gamma = g^2 N(0)$ and $N(0)$ being the bath DOS evaluated at the Fermi level. 
Note that in going to the Lindblad equation, the bath has been integrated out\cite{breuer2002theory}.

{We can use the Lindbladians to construct a Kraus map that approximately reconstruct the state at $t + \Delta t$ from the knowledge of the state at time $t$: $\rho(t+\Delta t) = \sum_i K^{\,}_i(t) \rho(t) K^{\dag}_i(t)$, where:
\begin{align}\label{eq:relation_Limbd_Kraus}
    K_0 &= \exp \left( i H(t) \Delta t \right)\sqrt{\mathbb{I}-\sum_i L^\dag_{i}(t) L^{\,}_i(t)\Delta t } \nonumber \\
    K_{i} & =\sqrt{\Delta t}L_i(t) \ \ \ \ \text{when }i = \ell\ge 1. 
\end{align}

From the knowledge of the Lindbladians one can use Eq.~(\ref{eq:relation_Limbd_Kraus}) to construct the Kraus operators that we defined in Eq.~(\ref{eq:kraus})
(see Ref.~\cite{del2020driven} for further details on the formal derivation).
}
\par The driven electric current is found from
\begin{equation}\label{eq:curr_1}
   J = \frac{\gamma_h}{\pi}\int dk \,\sin(k+\Omega \,t)n_k(t),
\end{equation}
where $n_k(t) = \mbox{Tr}\left[\rho_k(t)d^\dag_kd^{\,}_k\right]$.  By multiplying Eq.~(\ref{eq:me_lin}) times $d^\dag_kd^{\,}_k$ and computing the trace of both sides of the equation, we obtain the following differential equation for the momentum distribution function $\dot{n}_k = -2\Gamma(n_k(t) - n_F(\epsilon_k(t)))$. 
Solving yields 
\begin{equation}
 n_k(t) =n_k(0)e^{-2\Gamma t}+ 2\Gamma \int_0^t ds\,e^{-2\Gamma(t-s)}n_F[\epsilon(k + \Omega s)].
\end{equation}
Next, we introduce the gauge-invariant wave vector $k_m = k + \Omega t$ into the integral 
\begin{equation}
 n_k(t)=n_k(0)e^{-2\Gamma t}+ 2\Gamma \int_0^t ds\,e^{-2\Gamma(t-s)}n_F[\epsilon(k_m + \Omega (s-t))].
\end{equation}
Then for $t\to\infty$ at fixed $k_m$, we can perform the change of variables $y = k_m + \Omega(s-t)$ to yield the following formula for the momentum distribution function:
\begin{equation}
     n(k_m) = \frac{2\Gamma}{\Omega}\int_{-\infty}^{k_m}dy \,e^{\frac{2\Gamma}{\Omega}(y-k_m)}n_F[\epsilon(y)],
\end{equation}
which depends only on $k_m$.
This implies that for large enough time, we can reconstruct the full momentum distribution function by fixing the value of the crystalline momentum $k$ and letting $t$ run over one full Floquet period. In practice, we use the fact that the transients die off quickly to use the data averaged over one Floquet period to represent the long-time limit of the momentum distribution.
\subsection{Atomic Hubbard circuit}\label{sup:hubCirc}

We describe how to dissipatively prepare the thermal state of the atomic limit of the Hubbard model on a pair of qubits. Our Hamiltonian is given by
\begin{equation}
\mathcal{H}=Un_{\uparrow}n_{\downarrow}-\frac{\mu}{2} (n_{\uparrow}+n_{\downarrow})-\frac{B}{2}(n_{\uparrow}-n_{\downarrow})
\end{equation}
where $U$ is the on-site interaction strength, $\mu$ is the chemical potential, $B$ is the magnetic field in units of energy, and $n_{\alpha}$ is the occupation number operator for state $\ket{\alpha}$. Our desired mixed state is then $\tilde\rho=e^{-\beta \mathcal{H}}/\operatorname{Tr} e^{-\beta \mathcal{H}}$ where $\beta$ is the inverse temperature of the thermal state.

We prepare this state by iteratively applying a dissipative map, $\mathcal{E}$, to the qubits such that $\tilde\rho$ is the unique fixed point of $\mathcal{E}$. We define $\mathcal{E}$ by a set of Kraus operators $\{K_i\}$ such that
\begin{equation}\label{eq:HubKraMap}
    \rho_{i+1}=\mathcal{E}(\rho_i)=\sum_s K_s^{\,} \rho_i K_s^\dagger .
\end{equation}
 with $\sum_s K_s^\dagger K_s^{\,}=1$ and $\mathcal{E}(\tilde\rho)=\tilde\rho$.
Such a map may be realized by choosing Kraus operators which induce transitions between eigenstates of the Hamiltonian ($\ket{j}\!\bra{i}$) with fixed transition probabilities ($\gamma_{i,j}$). These transition probabilities define a detailed-balanced condition,
\begin{equation}\label{eq:detBal}
   \sum_j \gamma_{i,j}\bra{i}\!\rho\ket{i}=\sum_j \gamma_{i,j}\bra{j}\!\rho\ket{j},
\end{equation}
i.e. the density transitioning out of each state is exactly balanced by the density transitioning in. When this condition is met, $\rho$ is a fixed point of $\mathcal{E}$. To reduce the complexity of the final circuit, we take only a minimal set of these transitions that correspond to a cycle through the states as depicted in Fig.~\ref{fig:hubbard}(b).

Solving Eq.~(\ref{eq:detBal}) reveals $\gamma_{i,j}$ is proportional to the inverse Boltzmann factor for state $i$, i.e. $\gamma_{i,j}=\mathcal{N} \ e^{\beta \varepsilon_i}$ where $\varepsilon_i$ is the energy of state $\ket{i}$ given by $\mathcal{H}\ket{i}=\varepsilon_i\ket{i}$. $\mathcal{N}$ can be thought of as an effective time step or system/bath coupling strength. Because $\mathcal{E}$ must satisfy the normalization condition given in Eq.~(\ref{eq:HubKraMap}), we find $0\leq \gamma_{i,j}\leq 1$ which bounds the values $\mathcal{N}$ can take. To maximize convergence speed, we take $\mathcal{N}$ to be such that the largest $\gamma_{i,j}$ is 1.

Armed with the Kraus operators, we can build a circuit that implements $\mathcal{E}$ in a straightforward way. Note that the eigenstates of $\mathcal{H}$ are the computational-basis states, so we can transition between states with single-qubit rotations. To do this, we first map the state of the system onto an ancilla register. Then we rotate the system qubits, controlled on the ancilla register, to implement the transitions. A transition probability of $\gamma_{i,j}$ is achieved with rotation angle $\theta_{i,j}=2\sin^{-1}(\sqrt{\gamma_{i,j}})$. Finally the ancilla qubits are reset in preparation for the next step. This circuit is shown in Fig.~\ref{fig:hubbard}(a) and yields Kraus operators of the form $K_{i,j}=\sqrt{1-\gamma_{i,j}}\ket{i}\!\bra{i}-i\sqrt{\gamma_{i,j}}\ket{j}\!\bra{i}$, as desired.

\begin{figure}[htpb]
	\centering
	\includegraphics[width=\columnwidth]{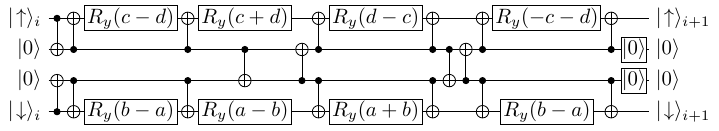}
	\caption{The circuit from Fig.~\ref{fig:hubbard}(a) transpiled into CNOT gates and single-qubit rotations suitable for nearest neighbor connectivity. We have adopted the shorthand $a = \Theta_{\uparrow\downarrow,\uparrow}/2$, $b = \Theta_{0,\downarrow}/2$, $c = \Theta_{\downarrow,\uparrow\downarrow}/2$, $d = \Theta_{\uparrow,0}/2$. We note that this circuit is the same as that shown in Fig.~\ref{fig:hubbard}(a) up to a unitary on the ancilla register before resetting and an $R_z(\pm\pi/2)$ gate before the first Trotter step and after the last Trotter step on both system qubits. These differences leave the dissipative map $\mathcal{E}$ unaffected with the initial $R_z(\pi/2)$ being absorbed into the initial state.}
	\label{fig:supHub}
\end{figure}

Ultimately this circuit needs to be transpiled down into IBM's native gates and conform to the qubit connectivity in a real device. This was done in part by hand and in part using the transpilation tools in Qiskit, the result of which is shown in Fig.~\ref{fig:supHub}.

\subsection{Error model}\label{sup:errMod}
The ideal curve for the steady state of the electron density, $n_k(t)$, is distorted by errors occurring in the quantum computer during the run. We can model this new distorted curve well by taking into account only two sources of error: 1) imperfect reset gates and 2) the $T_1$ decay of the system qubit.

We first discuss the reset gates. In IBM's hardware, a reset gate is performed by measuring the state of a qubit and then applying an $X$ gate if the qubit was measured to be in the $|1\rangle$ state and leaving it in the $|0\rangle$ state otherwise. However, even assuming a perfect $X$ gate, measurement fidelity is imperfect and so we expect an imperfect reset gate. Call the probability of measuring $m$, given the qubit is actually in $|m\rangle$ to be $p(m|m)\equiv p_m$. If the qubit is initially in state $\rho$ with $\langle 0|\rho |0\rangle=a_0$, then we model the state of the qubit after one reset gate, $\rho_1$, to be characterized by 
\begin{equation}
\langle 0|\rho_1|0\rangle=a_0 (p_0-p_1)+ p_1 \quad \text{and} \quad \langle 0|\rho_1|1\rangle=0.
\end{equation}
In IBM's hardware, we have $p_0\geq p_1$, so we can improve the fidelity of a reset operation by applying multiple reset gates in succession. Call the qubit state after $r$ reset gates $\rho_r$. Given the above model, we expect the probability of a successful reset to be
\begin{equation}\label{eq:resProb}
\langle 0|\rho_r|0\rangle=a_0 \left(p_0-p_1\right)^r+\frac{p_1 \left(1-\left(p_0-p_1\right)^r\right)}{1-p_0+p_1}.
\end{equation}
This simple model of reset infidelity appears sufficient to explain most of the observed results. We find $p_0=0.97$ and $p_1=0.91$ on average over all our runs.

Now, we discuss the $T_1$ decay. Because a reset gate is approximately an order of magnitude longer than the combined duration of the other operations in a single Trotter step, we assume all of amplitude damping of the system qubit occurs during the time it takes for the reset operation on the ancilla qubit. We can model the amplitude damping channel as performing the following map on a single qubit: $\rho\mapsto \sum_k M_k\rho M_k^\dagger\equiv \mathcal{E}_{x}(\rho)$ where $M_k\in\left\{\sqrt{1-e^{-rT}}|0\rangle\langle1|\right.,$ $\left .|0\rangle\langle0|+\sqrt{e^{-rT}}|1\rangle\langle1| \right\}$ are the Kraus operators for the amplitude damping channel, $T$ is the duration of a single reset gate in units of the $T_1$ time of the system qubit, and $r$ is the number of reset gates used in a reset operation. We find the average duration of a reset gate is approximately 6\% of the system qubit $T_1$ time.

Putting these together with the circuit shown in Fig.~\ref{fig:nk}(d) gives our error model. Call $\rho^S_s$ and $\rho^a_s$ the density matrix of the system and ancilla at Trotter step $s$ (or time $t=s\Delta t$), respectively. Because the error incurred during a single Trotter step propagates non-trivially to the next steps, we express our model below in terms of recursive formulae for the populations $n_s\equiv\bra{0}\rho_s^S\ket{0}$ and $a_s\equiv\bra{0}\rho_s^a\ket{0}$. We explicitly represent the unitary part of the circuit for step $s$ as a matrix $U_s$ so that applying the circuit to a generic two-qubit state $\rho$ gives $\rho\mapsto U_s\rho U_s^\dagger$. This yields

\begin{equation}
\rho_{s+1}=\rho_{s+1}^a\otimes \mathcal{E}_x\left(\text{Tr}_a\left[ U_s(\rho_s^a\otimes\rho_s^S) U_s^\dagger\right]\right),
\end{equation}
which leads to the following coupled recurrence relations
\begin{align}
n_{s+1}&=1-e^{-r T} \left(\left(2 a_s-1\right) \left(2 \Gamma  \Delta t  \left(n_s-n_F[\varepsilon_s]\right)-n_s\right)+a_s\right)\label{eq:errDen}\\
a_{s+1}&=\frac{a_s+2 \Gamma  \Delta t \left(2 a_s-1\right) \left(\left(n_s-1\right) n_F[\varepsilon_s]-n_s n_F[-\varepsilon_s ]\right)}{\left(p_0-p_1\right)^{-r}}\nonumber\\
&+\frac{p_1 \left(1-\left(p_0-p_1\right)^r\right)}{1-p_0+p_1}.
\end{align}
Here  $\varepsilon_s\equiv -2\cos(k+\Omega s \Delta t)$ is the dispersion relation,  $n_F[x]$ is the Fermi-Dirac distribution, and the angles at step $s$ are defined by
\begin{equation}
\theta_s=2 \sin ^{-1}\sqrt{\frac{2\Gamma  \Delta t}{e^{\beta  \varepsilon_s}+1}}\quad\text{and}\quad\phi_s=2 \sin ^{-1}\sqrt{\frac{2\Gamma  \Delta t}{e^{-\beta  \varepsilon_s}+1}}
\end{equation}

\begin{figure*}[t!]
	\centering
	\includegraphics[width=0.9\textwidth]{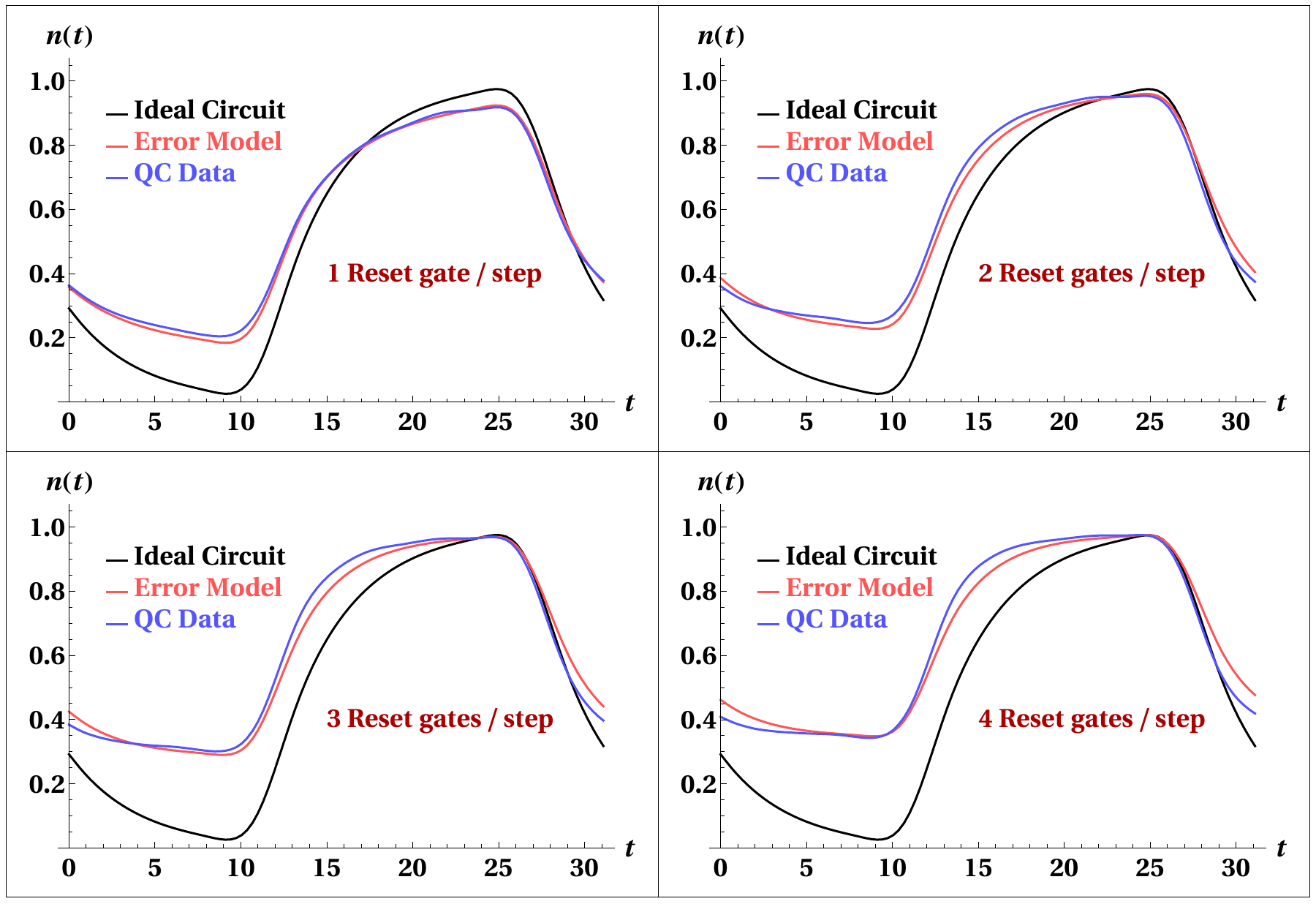}
	\caption{(Color online) Comparison of the ideal circuit, our error model and the actual data from the quantum computer for $r=1,2,3,$ and 4 reset gates per Trotter step. The ideal circuit and quantum computer data are the same as shown in Fig.~\ref{fig:nk}. The error model and data agree well overall and adequately explain the majority of the error observed.}
	\label{fig:supErr}
\end{figure*}

These recurrence relations can easily be solved numerically. We use an optimizer to find $T=0.06T_1, \ p_0=0.97,$ and $p_1=0.91$, which best match our data. The same values for $T, \ p_0,$ and $p_1$ are used for each $r=$ 1, 2, 3, and 4 iterated reset gates per reset operation, since the parameters should be approximately constant for a given pair of qubits on a given quantum device. A comparison of the error model vs. ideal circuit vs. actual quantum computer data is shown in Fig.~\ref{fig:supErr}.

This error model also gives insight into why our system, and dissipative systems in general, are robust against noise. Inspecting Eq.~(\ref{eq:errDen}) shows that $n_{s+1}$ (the electron density at step $s+1$) depends on $n_s$ as $e^{-rT}(2 a_s-1)(2\Gamma \Delta t-1)n_s$. Furthermore, we have $0\leq e^{-rT}, \ 2 a_s-1\leq 1$, and $0\leq2\Gamma \Delta t-1<1$ when $\Gamma>0$, i.e. when dissipation is non-zero. This tells us that the effect of an error at step $s$ on the density at step $s'>s$ is exponentially small in $s'-s$. This further suggests that if we have a sufficiently small base error rate, the compounding effect of these errors are offset by the dissipation, and this is why our error model does not need to take into account other errors, such as gate errors.

We can get an intuitive sense for how this works if we make the assumption that 
$a_s=a_0$ for all $s$, i.e. reset fidelity is independent of the qubit state. This is a good assumption when $p_0\approx p_1$ or $r\gg1$ since the reset fidelity depends on the qubit state through $\bra{0}\rho_a\ket{0}(p_0-p_1)^r$. This then recasts Eq.~(\ref{eq:errDen}) as 
\begin{equation}
    n_{s+1}=1-e^{-r T} \left(\left(2 a_0-1\right) \left(2 \Gamma  \Delta t  \left(n_s-n_F[\varepsilon_s]\right)-n_s\right)+a_0\right)
\end{equation}
for which Mathematica gives the analytic solution of the recurrence relation for $n_s$, $s\gg1$ as
\begin{align}\label{eq:errNs}
    \frac12&+\frac{1-e^{-r T}}{2(1-a_0 e^{-r T} (1-2 \Gamma  \Delta t))}\nonumber\\
    &-\frac{a_0}{e^{r T}} \Gamma  \Delta t\sum _{t=0}^{s-1} \tanh \left(\frac{\beta  \varepsilon _t}{2}\right) \left(\frac{a_0}{e^{r T}} (1-2 \Gamma \Delta t) \right)^{s-t-1}.
\end{align}
Plugging in $a_0=1$ and $T=0$ recovers the error free solution
\begin{equation}
    \frac12-\Gamma  \Delta t\sum _{t=0}^{s-1} \tanh \left(\frac{\beta  \varepsilon _t}{2}\right) \left((1-2 \Gamma \Delta t) \right)^{s-t-1}.
\end{equation}
These two solutions are qualitatively the same. They differ by a shift (second term in Eq~(\ref{eq:errNs})), a stretch of $a_0 e^{-r T}$ and deformation in shape equivalent to using $\Gamma'$ instead of $\Gamma$, such that $(1-2\Gamma'\Delta t)= a_0 e^{-r T}(1-2\Gamma\Delta t)$. This shows that the errors in the quantum hardware tend to perturb but not destroy the steady-state dynamics. We anticipate this result to hold true for many dissipative algorithms---as we can consider the error channels occurring inside the quantum computer to supplement the dissipative channels we are explicitly simulating.  As long as the error rates are low and/or our dissipation rate is high, we should expect to reach a nearly error-free steady state.

\subsection{Processing \texorpdfstring{$\boldsymbol{n_k(t)}$}{nk(t)}}\label{sup:proNkt}

After the transients have died (about 30 Trotter steps for the data in Fig.~\ref{fig:nk}), the population of the $|0\rangle$ state gives our raw data for $n_k(s\Delta t)$. We then correct this data for measurement errors using the built in tools in Qiskit Ignis~\cite{qiskit2024} to obtain the steady state data (e.g. data with green background in Fig.~\ref{fig:nk}(b)). We then discard these first Trotter steps as transient data. The remaining data is (ideally) periodic according to Floquet theory with a period of $\tau=2\pi/\Omega$, since our Hamiltonian is periodic with the same period (even though the field is DC, it drives periodic Bloch oscillations, making the system periodic). Because the step size $\Delta t$ is not always commensurate with the Floquet period, we use quadratic interpolation to generate a smooth curve, on a common time grid, and average together distinct periods to obtain an averaged curve for $n^{\text{ave}}_k(t)$ over a single period
\begin{equation}
n^{\text{ave}}_k(t)=\sum_{\ell=0}^{(t_{\text{max}}-t)/\tau} n_k(t+\ell\tau),
\end{equation}
where $t$ lies on the common time grid.
It is from this quantity that steady-state observables (e.g. DC current) can be calculated.

The constructed $n^{\text{ave}}_k(t)$ is distorted by errors occurring within the quantum computer, as detailed in the previous subsection. However, we can correct for the majority of this distortion. First, we know that $\int_0^\tau dt n_k(t)=0.5$, which allows us to center the momentum distribution curve about its true midpoint (at $n_k=0.5$). Second, we ran our circuits with $r=1,2,3$, and 4 native reset gates per Trotter step. We would like to be able to scale our results to an instantaneous application time for the resets, similar to Richardson extrapolation.  But, while the additional reset gates do boost the fidelity of the reset operation, they also add decoherence errors, because the gates are non-negligibly long compared to the $T_1$ time of the qubits, as described in the previous subsection. For this work, we find a negligible difference in reset fidelity between $r=2,3,4$. This then allows us to extrapolate to $r=0$, the limit of no $T_1$ decay. 

As can be seen in the previous subsection, the functional dependence of $n_k(t)$ on $T_1$ errors is complex and so we opt for using a simple quadratic extrapolation. Doing this tends to be reasonably effective in improving the data. However, this does nothing to correct reset infidelity. Furthermore, when we have a limited number of periods of steady-state data, or when $\Omega$ is large and the curves are quite flat, noise from counting statistics and other random errors inside the quantum computer make the extrapolation less reliable. This is especially true when computing the DC response, which is quite sensitive to the shape of $n_k(t)$. To mitigate these issues, we correct the amplitude by computing the theoretical maximum of $n_k(t)$ at zero temperature and then use that as an approximate scale factor. Starting with our master equation in terms of our gauge-invariant wavevector, $k_m=k+\Omega t$, we have
\begin{equation}
\dot n(k_m)=2\Gamma\left(\frac{1}{1+e^{2\beta\cos k_m}}-n(k_m)\right).
\end{equation}
\color{black}
Taking the zero temperature limit, $\beta\to\infty$, gives \color{black}
\begin{align}
\dot n_{k_m}(t)&=\Gamma\left(1+\operatorname{sgn}\left(\cos k_m\right)-n_{k_m}(t)\right)\nonumber\\
&\to
\begin{cases}
\cos k_m<0 & \dot n=-2\Gamma n\\
\cos k_m>0 & \dot n=2\Gamma(1-n).
\end{cases}
\end{align}
Since $0\leq n\leq 1$, the derivative changes sign when $\cos k_m=0$. \color{black}But the solution for $n_{k_m}(t)$ generically does not show full visibility, and instead ranges from a minimal value larger than 0 to a maximal value smaller than 1, as it evolves over time. \color{black} The differential equation can be immediately solved, and we find that 
\begin{equation}
n_\text{max}=\frac{1}{2}\left(1+\tanh\frac{\pi\Gamma}{\Omega}\right).
\end{equation}
We use this value to stretch the curves about $1/2$ (their midpoint), which gives our final curve. Despite the distortions remaining after the extrapolation due to reset infidelity, uncorrected $T_1$ effects and other sources of error, we generally find excellent agreement between the ideal result and our final post-processed curve.

\end{document}